\documentclass[preprint,prd,aps,showpacs,showkeys,nofootinbib]{revtex4}
\usepackage{amsmath,graphicx}
\usepackage{epstopdf}
\usepackage{subfigure}
\begin{document}
\title{$B \to X_{\mathrm{s}} l^{+} l^{-}$ in the $\mu$ from $\nu$ Supersymmetric Standard Model}

\author{
Xiao-Jie Hu$^{1,2,}$\footnote{xiaojiehu.research@outlook.com},
Hai-Bin Zhang$^{1,2,3,}$\footnote{hbzhang@hbu.edu.cn},
Tai-Fu Feng$^{1,2,3,}$\footnote{fengtf@hbu.edu.cn}
}

\affiliation{
$^1$Department of Physics, Hebei University, Baoding, 071002, China\\
$^2$Hebei Key Laboratory of High-precision Computation and Application of
Quantum Field Theory, Baoding, 071002, China\\
$^3$Hebei Research Center of the Basic Discipline for Computational Physics, Baoding, 071002, China
}

\begin{abstract}
We investigate the impact of new physics on the rare inclusive decay $B \to X_{\mathrm{s}} l^{+} l^{-}$ within the
framework of the $\mu$ from $\nu$ Supersymmetric Standard Model ($\mu\nu$SSM). The dominant contributions
to the relevant Wilson coefficients and their corresponding particles are identified and analyzed.
By performing a systematic scan over the relevant parameter space, we elucidate the underlying
physical mechanisms governing these dominant contributions and demonstrate their consistency
with the experimentally allowed regions. Experimental constraints from the decays  $\bar{B} \to X_{\mathrm{s}}\gamma$, $B_{\mathrm{s}}^{0} \to \mu^{+} \mu^{-}$, and the $125\,\text{GeV}$ SM-like Higgs boson are also incorporated. We perform a systematic interference decomposition of the Wilson-coefficient contributions to the forward-backward asymmetry, identifying the $C_7C_{10}$ and $C_9C_{10}$ interference terms as the dominant contributions governing its behavior in both the low- and high-$q^2$ regions.
\end{abstract}

\keywords{Rare decay, $B$ physics, $\mu\nu$SSM}
\pacs{12.60.Jv, 14.40.Nd}

\maketitle
\section{Introduction\label{sec1}}
Flavour-changing neutral current (FCNC) rare $B$ decays provide a promising window onto physics beyond the Standard Model (SM), as theoretical predictions for these processes suffer less from uncertainties associated with non-perturbative QCD effects. Charmless semileptonic $B$ decays have been extensively studied in the SM, with the dominant perturbative contributions evaluated in Ref.~\cite{Grinstein:NPB89}, two-loop QCD corrections discussed in Refs.~\cite{Misiak:PRD93,Buras:PRD95}, and the effects of $c\bar{c}$ resonances in $B \to X_{\mathrm{s}} l^{+} l^{-}$ presented in Refs.~\cite{Deshpande:PRD89,Lim:PLB89,Donnell:PRD91}.

In this work, we adopt the SM prediction for $B \to X_{\mathrm{s}} l^{+} l^{-}$ from Ref.~\cite{Huber:JHEP2024}. This calculation incorporates next-to-next-to-leading-order (NNLO) QCD corrections, next-to-leading-order (NLO) electroweak corrections, and power corrections in the heavy-quark expansion (HQE). Long-distance charmonium contributions are treated using the Kr\"uger-Sehgal (KS) approach, including $\frac{1}{m_c^2}$ corrections. The results are shown as
\begin{align}
&\mathrm{BR}(B \to X_{\mathrm{s}} l^{+} l^{-})_{q^2 \in [1,6]\, \mathrm{GeV}^2}^{\mathrm{SM}}
= (1.69 \pm 0.13) \times 10^{-6}, \nonumber \\
&\mathrm{BR}(B \to X_{\mathrm{s}} l^{+} l^{-})_{q^2 \in [14.4,25]\, \mathrm{GeV}^2}^{\mathrm{SM}}
= (3.04 \pm 0.69) \times 10^{-7},\quad (l = e, \mu).
\label{eq:SM-BR-BtoXsll}
\end{align}
The decay of $B \to X_{\mathrm{s}} l^{+} l^{-}$  has been further studied in the low $q^2$ region  $1\,\mathrm{GeV}^2 \leq q^2 \leq 6\,\mathrm{GeV}^2$ and the high $q^2$ region $14.4\,\mathrm{GeV}^2 \leq q^2 \leq 25\,\mathrm{GeV}^2$, where $q^2$ represents the dilepton invariant mass squared. The experimental averages of the branching ratios, based on the weighted averages from BaBar~\cite{BABAR:PRL04,Lees:PRL14} and Belle~\cite{BELLE:PRD05,BELLE:PRD16}, are as follows:
\begin{align}
&\mathrm{BR}(B \to X_{\mathrm{s}} l^{+} l^{-})_{q^2 \in [1,6]\, \mathrm{GeV}^2}^{\mathrm{exp}}
= (1.58 \pm 0.37) \times 10^{-6}, \nonumber \\
&\mathrm{BR}(B \to X_{\mathrm{s}} l^{+} l^{-})_{q^2 \in [14.4,25]\, \mathrm{GeV}^2}^{\mathrm{exp}}
= (4.8 \pm 1.0) \times 10^{-7},\quad (l = e, \mu),
\label{eq:EXP-BR-BtoXsll}
\end{align}
where the SM predictions are in agreement with the experimental results, indicating that precise measurements of $B$ decay processes impose tight constraints on new physics beyond the SM. In addition to the branching ratios, the forward-backward asymmetries can also probe new physics scenarios. The updated experimental data on $B \to X_{\mathrm{s}} l^{+} l^{-}$ in the low-and high-$q^2$  regions are summarized as~\cite{BELLE:PRD16}
\begin{align}
&A_{\mathrm{FB}}(B \to X_{\mathrm{s}} l^{+} l^{-})_{q^2 \in [1,6]\, \mathrm{GeV}^2}^{\mathrm{exp}}
= 0.30 \pm 0.24 \pm 0.04, \nonumber \\
&A_{\mathrm{FB}}(B \to X_{\mathrm{s}} l^{+} l^{-})_{q^2 \in [14.4,25]\, \mathrm{GeV}^2}^{\mathrm{exp}}
= 0.28 \pm 0.15 \pm 0.02.
\label{eq:exp-AFB0}
\end{align}
The corresponding SM predictions are~\cite{AFB:PRD02}
\begin{align}
&A_{\mathrm{FB}}(B \to X_{\mathrm{s}} l^{+} l^{-})_{q^2 \in [1,6]\, \mathrm{GeV}^2}^{\mathrm{SM}}
= -0.07 \pm 0.04, \nonumber \\
&A_{\mathrm{FB}}(B \to X_{\mathrm{s}} l^{+} l^{-})_{q^2 \in [14.4,25]\, \mathrm{GeV}^2}^{\mathrm{SM}}
= 0.40 \pm 0.04.
\label{eq:SM-AFB0}
\end{align}
Owing to current experimental uncertainties, the forward-backward asymmetries cannot yet be used to rigorously test the Standard Model. With improved precision and increased statistics in future measurements, these observables are expected to impose more restrictive constraints on the parameter space of new physics.

Among extensions of the Standard Model, supersymmetry is one of the most promising candidates. The constraints on new physics beyond the SM have been extensively studied in the literature. The branching ratio for the rare inclusive decay $\bar{B} \to X_{\mathrm{s}}\gamma$ has been calculated in Refs.~\cite{BXR1,BXR2Hewett93}, with QCD corrections presented in Refs.~\cite{Ciafaloni:NPB524,Ciuchini:NPB527,Borzumati:PRD58,Borzumati:PRD62}. Refs.~\cite{He:1988PRD38,Skiba:1993NPB404,Cho:PR54,Choudhury:1998PLB451,Huang:2000PRD63,Feng:2016PRD94,Feng:2016IJMP31,Yang:EPJC18} calculated the inclusive decay $B \to X_{\mathrm{s}} l^{+} l^{-}$ within the framework of the Two-Higgs Doublet Model (THDM) and the Minimal Supersymmetric Standard Model (MSSM). In addition, the measured branching ratios of $\bar{B} \to X_{\mathrm{s}}\gamma$ and $B \to X_{\mathrm{s}} l^{+} l^{-}$ have been used to constrain relevant parameters, as discussed in Refs.~\cite{Barbieri:PLB309,Ali:ZPC67,Shaw:PRD69}. The authors of Refs.~\cite{Ali:PRD61,Huber:JHEP10SMP} presented predictions for the branching ratios and forward-backward asymmetries in $B \to (K,K^*)l^+l^-$ and $B \to X_{\mathrm{s}} l^{+} l^{-}$ decays. While the effects of SM extensions on $B$ meson decays have been widely reviewed in Refs.~\cite{Masiero:172,Masieroz:404}, this study specifically focuses on the constraints on the parameter space imposed by rare decay channels. Adopting the mass insertion approximation (MIA) method, Refs.~\cite{Lunghi:NPB568,Xiao:CTP46} analyzed the dependence of relevant mass-insertion parameters and derived constraints on the model parameters from the processes $B \to X_{\mathrm{s}} l^{+} l^{-}$.

In contrast to the Minimal Supersymmetric Standard Model~\cite{Haber:85,Rosiek:90}, the $\mu\nu$SSM~\cite{Fogliani:PRL06,Escudero:JHEP08,Fidalgo:JHEP10} addresses the $\mu$ problem of MSSM~\cite{Nilles:PLB84} via the R-parity breaking coupling  in the  superpotential. When the electroweak symmetry is broken, the $\mu$ term is spontaneously generated by the nonzero right-handed sneutrino vacuum expectation values (VEVs), $\mu = \lambda_i \langle \tilde{\nu}_i^c \rangle$. Furthermore, a TeV-scale seesaw mechanism can generate three tiny neutrino masses at tree level. The neutrino mass and mixing patterns in the $\mu\nu$SSM have been systematically studied in our previous work~\cite{neutrino-mass13}, and the parameter regions adopted in the present analysis are consistent with those constraints.

In our previous work, we studied the decay  $\bar{B} \to X_{\mathrm{s}}\gamma$ and $B_{\mathrm{s}}^{0} \to \mu^{+} \mu^{-}$ in the $\mu\nu$SSM~\cite{Zhang:MPLA29,MengKK:PRD111}. In this paper, we investigate the FCNC process $B \to X_{\mathrm{s}} l^{+} l^{-}$ in the model. The paper is organized as follows. In Section II, we briefly introduce the $\mu\nu$SSM and present the effective Hamiltonian for $B \to X_{\mathrm{s}} l^{+} l^{-}$ in Section III. The formulas for the branching ratios and forward-backward asymmetries are shown in Section IV. The numerical analysis is given in Section V. Additional constraints are discussed in Section VI, and Section VII provides a summary.

\section{the $\mu\nu$SSM\label{sec2}}
Compared to the superfield of the MSSM, the $\mu\nu$SSM contains three singlet neutrino superfields. The superpotential of the $\mu\nu$SSM is expressed as~\cite{Fogliani:PRL06}
\begin{equation}
\begin{aligned}
W = & \epsilon_{ab} \big( Y_{u_{ij}} \hat{H}_u^b \hat{Q}_i^a \hat{u}_j^c
       + Y_{d_{ij}} \hat{H}_d^a \hat{Q}_i^b \hat{d}_j^c
       + Y_{e_{ij}} \hat{H}_d^a \hat{L}_i^b \hat{e}_j^c
       + Y_{\nu_{ij}} \hat{H}_u^b \hat{L}_i^a \hat{\nu}_j^c \big) \\
    & - \epsilon_{ab} \lambda_i \hat{\nu}_i^c \hat{H}_d^a \hat{H}_u^b
    + \frac{1}{3} \kappa_{ijk} \hat{\nu}_i^c \hat{\nu}_j^c \hat{\nu}_k^c \ .
\end{aligned}
\label{eq:superpotential}
\end{equation}
Here, $a,b=1,2$ represent the $SU(2)$ indices, with antisymmetric tensor $\epsilon_{ab}$, and $i,j,k = 1,2,3$ are generation indices. $\hat{H}_d^T = \left( \hat{H}_d^0, \hat{H}_d^- \right)$, $\hat{H}_u^T = \left( \hat{H}_u^+, \hat{H}_u^0 \right)$, $\hat{Q}_i^T = \left( \hat{u}_i, \hat{d}_i \right)$ and  $\hat{L}_i^T = \left( \hat{\nu}_i, \hat{e}_i \right)$ represent $SU(2)$  doublet superfields, and $\hat{d}_j^c$, $\hat{u}_j^c$, $\hat{e}_j^c$ denote the singlet down-type, up-type and charged lepton superfields, respectively. Moreover, $Y$, $\kappa$ and $\lambda$ are dimensionless matrices, a totally symmetric tensor, and a vector, respectively. The last two terms in Eq.~\eqref{eq:superpotential} clearly violate lepton number and R-parity.
After EWSB, the right-handed sneutrino expectation values (VEVs) can spontaneously generate the effective bilinear terms $\epsilon_{ab}\mu\hat{H}_d^a \hat{H}_u^b$, where $\mu$ is defined as $\mu = \lambda_i \langle \tilde{\nu}_i^c \rangle$. In addition, this electroweak symmetry breaking mechanism leads to the generation of the effective Majorana masses for neutrinos through the last term in Eq.~\eqref{eq:superpotential}. In this paper, the summation convention is implied on repeated indices.

Once the electroweak symmetry is spontaneously broken, the neutral scalars generally develop VEVs as follows:
\begin{align}
\langle H_d^0 \rangle = \upsilon_d , \qquad \langle H_u^0 \rangle = \upsilon_u , \qquad
	\langle \tilde \nu_i \rangle = \upsilon_{\nu_i} , \qquad \langle \tilde \nu_i^c \rangle = \upsilon_{\nu_i^c}.\label{VEVs}
\end{align}
Thus, the neutral scalars can be defined as
\begin{align}
&&H_d^0=\frac{h_d + i P_d}{\sqrt{2}} + \upsilon_d, \qquad\; \tilde \nu_i = \frac{(\tilde \nu_i)^\Re + i (\tilde \nu_i)^\Im}{\sqrt{2}} + \upsilon_{\nu_i},  \nonumber\\
&&H_u^0=\frac{h_u + i P_u}{\sqrt{2}} + \upsilon_u, \qquad \tilde \nu_i^c = \frac{(\tilde \nu_i^c)^\Re + i (\tilde \nu_i^c)^\Im}{\sqrt{2}} + \upsilon_{\nu_i^c},
\end{align}
and
\begin{eqnarray}
	\tan\beta=\frac{\upsilon_u}{\sqrt{\upsilon_d^2+\upsilon_{\nu_i}\upsilon_{\nu_i}}}.
\end{eqnarray}
where $\upsilon_{\nu_i}$ and $\upsilon_{\nu_i^c}$ represent the VEVs of left-handed and right-handed sneutrino superfields, respectively. Given that \( \upsilon_{\nu_i} \ll \upsilon_d, \upsilon_u \), we can define the value of \( \tan \beta \) as usual, where \( \tan \beta = \frac{\upsilon_u}{\upsilon_d} \).

In the $\mu\nu$SSM, the general soft SUSY-breaking terms are written as
\begin{align}
- \mathcal{L}_{\mathrm{soft}} =&\ m_{\tilde{Q}_{ij}}^2 \tilde{Q}{_i^{a*}} \tilde{Q}_j^a
+ m_{\tilde{u}_{ij}^c}^2 \tilde{u}{_i^{c*}} \tilde{u}_j^c
+ m_{\tilde{d}_{ij}^c}^2 \tilde{d}{_i^{c*}} \tilde{d}_j^c
+ m_{\tilde{L}_{ij}}^2 \tilde{L}_i^{a*} \tilde{L}_j^a \nonumber\\
&\ + m_{\tilde{e}_{ij}^c}^2 \tilde{e}{_i^{c*}} \tilde{e}_j^c
+ m_{H_d}^2 H_d^{a*} H_d^a
+ m_{H_u}^2 H{_u^{a*}} H_u^a
+ m_{\tilde{\nu}_{ij}^c}^2 \tilde{\nu}{_i^{c*}} \tilde{\nu}_j^c \nonumber\\
&\ + \epsilon_{ab} \bigl[ (A_u Y_u)_{ij} H_u^b \tilde{Q}_i^a \tilde{u}_j^c
+ (A_d Y_d)_{ij} H_d^a \tilde{Q}_i^b \tilde{d}_j^c + (A_e Y_e)_{ij} H_d^a \tilde{L}_i^b \tilde{e}_j^c + \mathrm{H.c.} \bigr] \nonumber\\
&\ + \bigl[ \epsilon_{ab} (A_\nu Y_\nu)_{ij} H_u^b \tilde{L}_i^a \tilde{\nu}_j^c
- \epsilon_{ab} (A_\lambda \lambda)_i \tilde{\nu}_i^c H_d^a H_u^b+ \frac{1}{3} (A_\kappa \kappa)_{ijk} \tilde{\nu}_i^c \tilde{\nu}_j^c \tilde{\nu}_k^c + \mathrm{H.c.} \bigr] \nonumber\\
&\ - \frac{1}{2} \bigl( M_3 \tilde{\lambda}_3 \tilde{\lambda}_3
+ M_2 \tilde{\lambda}_2 \tilde{\lambda}_2
+ M_1 \tilde{\lambda}_1 \tilde{\lambda}_1 + \mathrm{H.c.} \bigr).
\label{soft-term12}
\end{align}
Here, the first eight terms are mass squared terms for sleptons, Higgs scalars and squarks. In the last three terms, the Majorana mass corresponding to gauginos $\tilde{\lambda}_3$, $\tilde{\lambda}_2$, and $\tilde{\lambda}_1$ is denoted by $M_3$, $M_2$, and $M_1$, respectively. The remaining terms consist of the trilinear scalar couplings. In addition to the terms from $\mathcal{L}_{\mathrm{soft}}$, $D$-terms and $F$-terms  also contribute to the tree-level scalar potential~\cite{Escudero:JHEP08}.

The \(8 \times 8\) mass matrix \(M^2_{S^{\pm}}\) for the charged scalar fields produces the massless, unphysical Goldstone bosons \(G^{\pm}\), as described in Refs.~\cite{Zhang:JHEP69,Zhang:LFV14,Zhang:PRD14}:
\begin{eqnarray}
	G^{\pm} = \frac{1}{\sqrt{\upsilon_d^2+\upsilon_u^2+\upsilon_{\nu_i} \upsilon_{\nu_i}}}
\bigl(\upsilon_d H_d^{\pm} - \upsilon_u H_u^{\pm} - \upsilon_{\nu_i} \tilde{e}_{L_i}^{\pm} \bigr).
\end{eqnarray}
In the unitary gauge, where the $W$ boson absorbs the Goldstone bosons \(G^{\pm}\), and its squared mass is given by
\begin{eqnarray}
m_W^2 &=& \frac{e^2}{2 s_W^2}
\bigl( \upsilon_u^2 + \upsilon_d^2 + \upsilon_{\nu_i} \upsilon_{\nu_i} \bigr)
\end{eqnarray}
Here, $e$ represents the electromagnetic coupling constant, and $s_W$ is defined as $\sin\theta_W$, with $\theta_W$ denoting the Weinberg angle.

\section{Effective Hamilton for  $B \to X_{\mathrm{s}} l^{+} l^{-}$ $(l = e, \mu, \tau)$\label{sec3}}
The effective Hamiltonian for the rare decay $b \to s l^{+} l^{-}$ transition at the hadronic scale reads
\begin{align}
\mathcal{H}_{\mathrm{eff}} = -\frac{4G_{\mathrm{F}}}{\sqrt{2}} V_{ts}^* V_{tb} \Biggl[
&C_1 \mathcal{O}_1^c + C_2 \mathcal{O}_2^c + \sum_{i=3}^{6} C_i \mathcal{O}_i
+ \sum_{i=7}^{10} (C_i \mathcal{O}_i + C_i' \mathcal{O}_i') \nonumber \\
&+ \sum_{i=S,P} (C_i \mathcal{O}_i + C_i' \mathcal{O}_i') \Biggr],
\label{eq:heff}
\end{align}
where the effective operators  \(\mathcal{O}_i (i = 1, 2, \ldots, 10, S, P)\) and \(\mathcal{O}_i' (i = 7, 8, \ldots, 10, S, P)\) can be expressed as~\cite{Grigjanis22,Buchalla68,Lin24,Yang59,Goertz84}
\begin{equation}
\begin{alignedat}{2}
& \hspace{2em} \mathcal{O}_1^c = (\bar{s}_L \gamma_\mu T^a c_L)(\bar{c}_L \gamma^\mu T^a b_L), &\quad&
\mathcal{O}_2^c = (\bar{s}_L \gamma_\mu c_L)(\bar{c}_L \gamma^\mu b_L), \\
& \hspace{2em} \mathcal{O}_3 = (\bar{s}_L \gamma_\mu b_L) \sum_q (\bar{q} \gamma^\mu q), &\quad&
\mathcal{O}_4 = (\bar{s}_L \gamma_\mu T^a b_L) \sum_q (\bar{q} \gamma^\mu T^a q), \\
& \hspace{2em} \mathcal{O}_5 = (\bar{s}_L \gamma_\mu \gamma_\nu \gamma_\rho b_L) \sum_q (\bar{q} \gamma^\mu \gamma^\nu \gamma^\rho q), &\quad&
\mathcal{O}_6 = (\bar{s}_L \gamma_\mu \gamma_\nu \gamma_\rho T^a b_L) \sum_q (\bar{q} \gamma^\mu \gamma^\nu \gamma^\rho T^a q), \\
& \hspace{2em} \mathcal{O}_7 = \frac{e}{16\pi^2} m_b (\bar{s}_L \sigma_{\mu\nu} b_R) F^{\mu\nu}, &\quad&
\mathcal{O}_7' = \frac{e}{16\pi^2} m_b (\bar{s}_R \sigma_{\mu\nu} b_L) F^{\mu\nu}, \\
& \hspace{2em} \mathcal{O}_8 = \frac{g_{\mathrm{s}}}{16\pi^2} m_b (\bar{s}_L \sigma_{\mu\nu} T^a b_R) G^{a,\mu\nu}, &\quad&
\mathcal{O}_8' = \frac{g_{\mathrm{s}}}{16\pi^2} m_b (\bar{s}_R \sigma_{\mu\nu} T^a b_L) G^{a,\mu\nu}, \\
& \hspace{2em} \mathcal{O}_9 = \frac{e^2}{g_{\mathrm{s}}^2} (\bar{s}_L \gamma_\mu b_L) \bar{l} \gamma^\mu l, &\quad&
\mathcal{O}_9' = \frac{e^2}{g_{\mathrm{s}}^2} (\bar{s}_R \gamma_\mu b_R) \bar{l} \gamma^\mu l, \\
& \hspace{2em} \mathcal{O}_{10} = \frac{e^2}{g_{\mathrm{s}}^2} (\bar{s}_L \gamma_\mu b_L) \bar{l} \gamma^\mu \gamma_5 l, &\quad&
\mathcal{O}_{10}' = \frac{e^2}{g_{\mathrm{s}}^2} (\bar{s}_R \gamma_\mu b_R) \bar{l} \gamma^\mu \gamma_5 l, \\
& \hspace{2em} \mathcal{O}_S = \frac{e^2}{16\pi^2} m_b (\bar{s}_L b_R) \bar{l} l, &\quad&
\mathcal{O}_S' = \frac{e^2}{16\pi^2} m_b (\bar{s}_R b_L) \bar{l} l, \\
& \hspace{2em} \mathcal{O}_P = \frac{e^2}{16\pi^2} m_b (\bar{s}_L b_R) \bar{l} \gamma_5 l, &\quad&
\mathcal{O}_P' = \frac{e^2}{16\pi^2} m_b (\bar{s}_R b_L) \bar{l} \gamma_5 l,
\end{alignedat}
\label{eq:operators}
\end{equation}
Here, $g_{\mathrm{s}}$ denotes the strong coupling constant, $F^{\mu\nu}$ is the electromagnetic field strength tensor, $G^{\mu\nu}$ represents the gluon field strength tensor, and $T^a$ (with $a = 1,\dots,8$) are the generators of the $SU(3)$ group.
\begin{figure}[t]
\centering

\subfigure[]{
  \includegraphics[width=0.31\textwidth]{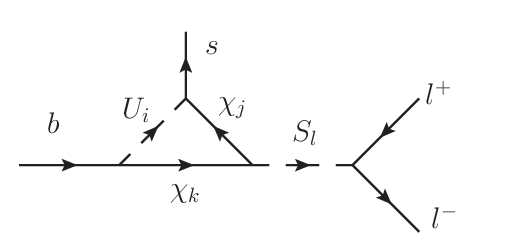}
}
\hfill
\subfigure[]{
  \includegraphics[width=0.31\textwidth]{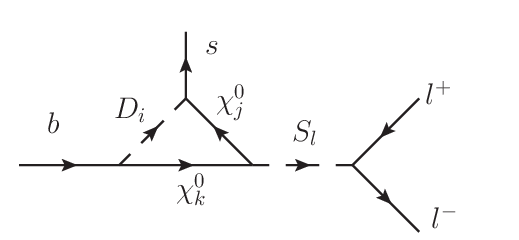}
}
\hfill
\subfigure[]{
  \includegraphics[width=0.31\textwidth]{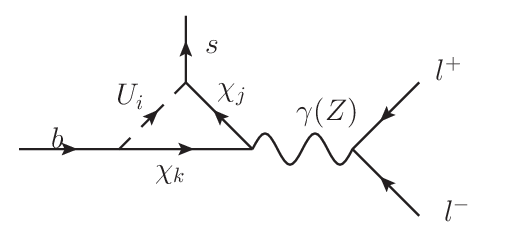}
}

\vspace{2pt}

\subfigure[]{
  \includegraphics[width=0.31\textwidth]{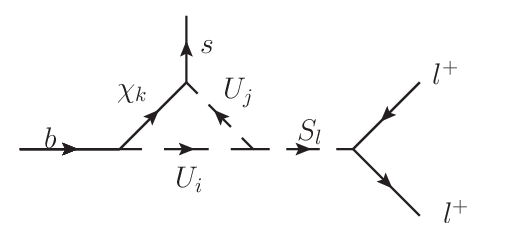}
}
\hfill
\subfigure[]{
  \includegraphics[width=0.31\textwidth]{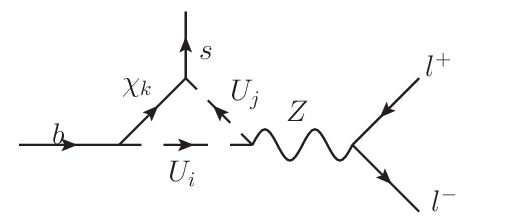}
}
\hfill
\subfigure[]{
  \includegraphics[width=0.31\textwidth]{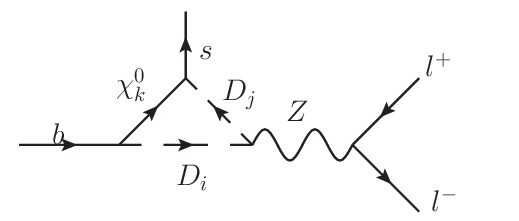}
}

\vspace{2pt}

\subfigure[]{
  \includegraphics[width=0.31\textwidth]{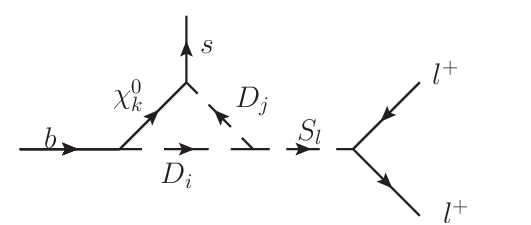}
}
\hfill
\subfigure[]{
  \includegraphics[width=0.31\textwidth]{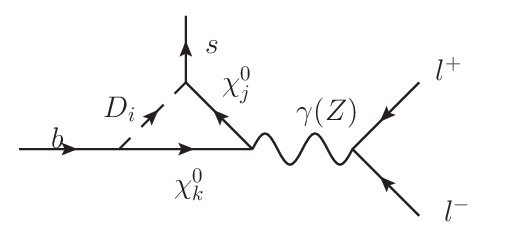}
}
\hfill
\subfigure[]{
  \includegraphics[width=0.24\textwidth]{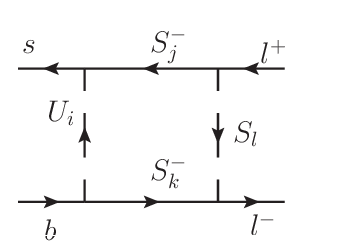}}

\vspace{3pt}

\centering
\subfigure[]{\includegraphics[width=0.24\textwidth]{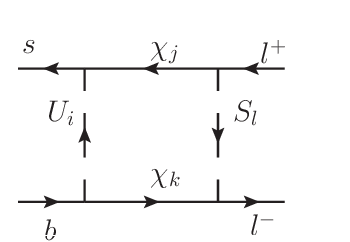}}
\hspace{0.15\textwidth}
\subfigure[]{\includegraphics[width=0.24\textwidth]{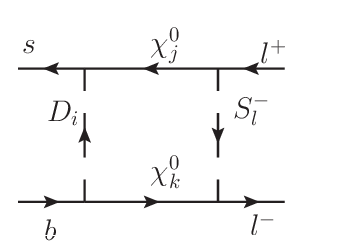}}

\caption{
Feynman diagrams contributing to the $b \to s l^{+} l^{-}$ transition in the $\mu\nu$SSM.}
\label{fig:feynman}
\end{figure}

Given the many unknown parameters that affect quark mass eigenstates and mixing matrices, which subsequently determine flavor-changing vertices, we adopt the mass insertion approximation (MIA) to evaluate the supersymmetric contributions to flavor-changing neutral current (FCNC) processes. In this framework, a basis is chosen for the fermion and sfermion states such that all couplings to neutral gaugino fields are flavor-diagonal. Flavor-changing effects arise solely from the off-diagonal elements of the squark propagators, allowing the pattern of flavor violation to be expressed systematically in terms of mass insertion parameters. The relevant flavor-violating parameters are defined as follows:
\begin{equation}
\tilde{M}^2_{q_{ij}} \simeq M^2_{\tilde{q}} (1 + \delta_{ij}^{AB})
\end{equation}
where $M_{\tilde{q}}$ is the common squark mass, $AB =LL, LR, RR $ denote the chirality, $i,j = 1,2,3$ are the generation indices, and $q = u,d$ labels up- and down-type squarks. Since we focus on the $b \to s$ decay process, only the 2nd and 3rd generation mixing terms are considered. The relevant flavor-violating parameters are assumed to satisfy:
\begin{equation}
\delta_{23}^{AB} \equiv \delta_{23}^{U,AB} = \delta_{23}^{D,AB}, \quad AB = LL, LR, RR.
\end{equation}
The flavor-violating parameters $\delta_{23}^{AB}$ originate from  trilinear scalar couplings and flavor-violating scalar mass terms in the soft breaking terms. For simplicity, we assume that the second- and third-generation flavor mass insertions are equal in both the up- and down-type squark sectors. The primary Feynman diagrams contributing to the process $b \to s l^{+} l^{-}$ in the $\mu\nu$SSM are shown in Fig.~\ref{fig:feynman}. Where ${\chi}_k$ ($k = 1,\dots,5$) denote charginos, ${\chi}_j^0$ ($j = 1,\dots,10$) denote neutralinos, $S_l$ ($l = 1,\dots,8$) denote neutral scalars, ${U}_i$ ($i = 1,\dots,6$) denote up-type squarks, ${D}_i$ ($i = 1,\dots,6$) denote down-type squarks, $S_i^-$ ($l = 2,\dots,8$) denote charged scalars, and $\tilde{g}$ denote gluino. Due to the large number of possible mass insertions, they are not explicitly marked on the diagrams, but all insertions are included in the calculation of the Wilson coefficients. At the electroweak energy scale $\mu_{\mathrm{EW}}$, the relevant Wilson coefficients are separated into
\begin{align}
& C_{7,NP}(\mu_{EW}) = C_{7,S^{\pm}}(\mu_{EW}) + C_{7,\chi^\pm}(\mu_{EW}) + C_{7,\chi^0}(\mu_{EW}) + C_{7,\tilde{g}}(\mu_{EW}) \, , \nonumber \\
& C_{8,NP}(\mu_{EW}) = C_{8,S^{\pm}}(\mu_{EW}) + C_{8,\chi^\pm}(\mu_{EW}) + C_{8,\chi^0}(\mu_{EW}) + C_{8,\tilde{g}}(\mu_{EW}) \, , \nonumber \\
& C_{9,NP}(\mu_{EW}) = C_{9,NP}^\gamma(\mu_{EW}) + C_{9,NP}^Z(\mu_{EW}) + C_{9,NP}^{box}(\mu_{EW}) \, , \nonumber \\
& C_{10,NP}(\mu_{EW}) = C_{10,NP}^\gamma(\mu_{EW}) + C_{10,NP}^Z(\mu_{EW}) + C_{10,NP}^{box}(\mu_{EW}) \, , \nonumber \\
& C_{S,NP}(\mu_{EW}) = \sum_{l=1}^2 C_{S,NP}^{S_l}(\mu_{EW}) + C_{S,NP}^{box}(\mu_{EW}) \, , \nonumber \\
& C_{P,NP}(\mu_{EW}) = C_{P,NP}^{S_l}(\mu_{EW}) + C_{P,NP}^{box}(\mu_{EW}) \, .
\label{eq:WC12}
\end{align}
Here, the subscript NP denotes new physics contributions. The symbols $\gamma$, $Z$, $S_l$, and box indicate that corrections to the respective Wilson coefficients arise from $\gamma$-, $Z$-, $S_l$- penguins, and box diagrams. For clarity, we further decompose these contributions as follows:
\begin{align}
& C_{9,NP}^\gamma(\mu_{EW}) = C_{9,S^{\pm}}^\gamma(\mu_{EW}) + C_{9,\chi^\pm}^\gamma(\mu_{EW}) + C_{9,\chi^0}^\gamma(\mu_{EW}) + C_{9,\tilde{g}}^\gamma(\mu_{EW}) \, , \nonumber \\
& C_{9,NP}^Z(\mu_{EW}) = (4s_W^2 - 1)C_{10,NP}^Z(\mu_{EW}) \, , \nonumber \\
& C_{10,NP}^Z(\mu_{EW}) = C_{10,S^{\pm}}^Z(\mu_{EW}) + C_{10,\chi^\pm}^Z(\mu_{EW}) + C_{10,\chi^0}^Z(\mu_{EW}) + C_{10,\tilde{g}}^Z(\mu_{EW}) \, , \nonumber \\
& C_{S,NP}^{S_l}(\mu_{EW}) = C_{S,S^{\pm}}^{S_l}(\mu_{EW}) + C_{S,\chi^\pm}^{S_l}(\mu_{EW}) + C_{S,\chi^0}^{S_l}(\mu_{EW}) + C_{S,\tilde{g}}^{S_l}(\mu_{EW}) \, , \nonumber \\
& C_{P,NP}^{S_l}(\mu_{EW}) = C_{P,S^{\pm}}^{S_l}(\mu_{EW}) + C_{P,\chi^\pm}^{S_l}(\mu_{EW}) + C_{P,\chi^0}^{S_l}(\mu_{EW}) + C_{P,\tilde{g}}^{S_l}(\mu_{EW}) \, , \nonumber \\
& C_{9,NP}^{box}(\mu_{EW}) = C_{9,S^{\pm}}^{box}(\mu_{EW}) + C_{9,\chi^\pm}^{box}(\mu_{EW}) + C_{9,\chi^0}^{box}(\mu_{EW}) + C_{9,\tilde{g}}^{box}(\mu_{EW}) \, , \nonumber \\
& C_{10,NP}^{box}(\mu_{EW}) = C_{10,S^{\pm}}^{box}(\mu_{EW}) + C_{10,\chi^\pm}^{box}(\mu_{EW}) + C_{10,\chi^0}^{box}(\mu_{EW}) + C_{10,\tilde{g}}^{box}(\mu_{EW}) \, , \nonumber \\
& C_{S,NP}^{box}(\mu_{EW}) = C_{S,S^{\pm}}^{box}(\mu_{EW}) + C_{S,\chi^\pm}^{box}(\mu_{EW}) + C_{S,\chi^0}^{box}(\mu_{EW}) + C_{S,\tilde{g}}^{box}(\mu_{EW}) \, , \nonumber \\
& C_{P,NP}^{box}(\mu_{EW}) = C_{P,S^{\pm}}^{box}(\mu_{EW}) + C_{P,\chi^\pm}^{box}(\mu_{EW}) + C_{P,\chi^0}^{box}(\mu_{EW}) + C_{P,\tilde{g}}^{box}(\mu_{EW}) \,.
\end{align}
Here, the detailed expressions of these Wilson coefficients are provided in the Supplemental Material. These Wilson coefficients serve as the basis for our calculations, and the dominant contributions will be identified in the subsequent numerical analysis. According to Eq.~\eqref{RQ}, the coefficients $C_7$ and $C_9$ are always correlated with other $C_i$ in the form of definite linear combinations. To conveniently obtain the hadronic scale matrix elements, the effective coefficients are written as~\cite{Buras:NPB424}
\begin{align}
C_7^{eff} &= \frac{4\pi}{\alpha_s} C_7 - \frac{1}{3} C_3 - \frac{4}{9} C_4 - \frac{20}{3} C_5 - \frac{80}{9} C_6 \, , \nonumber \\
C_8^{eff} &= \frac{4\pi}{\alpha_s} C_8 + C_3 - \frac{1}{6} C_4 + 20 C_5 - \frac{10}{3} C_6 \, , \nonumber \\
C_9^{eff} &= C_9^{eff,SM} + Y(q^{2}) \, , \nonumber \\
C_{10}^{eff} &= \frac{4\pi}{\alpha_s} C_{10} \, , \quad
C_{7,8,9,10}^{\prime eff} = \frac{4\pi}{\alpha_s} C_{7,8,9,10}^\prime \, ,
\end{align}
\label{eq:19}
where the functions  $Y(q^{2})$  are given by
\begin{align}
\mathrm{Y}(q^{2}) &= \mathrm{h}(q^{2}, m_{c}) \Big( \tfrac{4}{3} C_{1} + C_{2} + 6 C_{3} + 60 C_{5} \Big)
        - \tfrac{1}{2} \mathrm{h}(q^{2}, m_{b}) \Big( 7 C_{3} + \tfrac{4}{3} C_{4} + 76 C_{5} + \tfrac{64}{3} C_{6} \Big) \nonumber \\
       &\quad - \tfrac{1}{2} \mathrm{h}(q^{2}, 0) \Big( C_{3} + \tfrac{4}{3} C_{4} + 16 C_{5} + \tfrac{64}{3} C_{6} \Big)
        + \tfrac{4}{3} C_{3} + \tfrac{64}{9} C_{5} + \tfrac{64}{27} C_{6}.
\end{align}
The function
\begin{equation}
\mathrm{h}(q^{2}, m_{q}) = -\frac{4}{9} \Big( \ln\frac{m_{q}^{2}}{\mu^{2}} - \frac{2}{3} - z \Big)
              - \frac{4}{9} (2+z) \sqrt{\lvert z-1 \rvert}\,
\begin{cases}
\arctan\dfrac{1}{\sqrt{z-1}}, & z > 1, \\[6pt]
\ln\dfrac{1+\sqrt{1-z}}{\sqrt{z}} - \frac{\mathrm{i}\pi}{2}, & z \le 1,
\end{cases}
\label{eq:hfunc}
\end{equation}
with $z = 4m_q^2/q^2$.
\begin{table}[t]
\centering
\caption{Effective SM Wilson coefficients evaluated at the hadronic
scale $\mu = m_b \simeq 4.8~\mathrm{GeV}$ with NNLL accuracy.}
\begin{tabular}{|c|c|c|c|}
\hline\hline
$C_7^{\mathrm{eff,SM}}$
& $C_8^{\mathrm{eff,SM}}$
& $C_9^{\mathrm{eff,SM}}$
& $C_{10}^{\mathrm{eff,SM}}$ \\
\hline
$-0.304$ & $-0.167$ & $4.211$ & $-4.103$ \\
\hline\hline
\end{tabular}
\label{tab:SMWC1}
\end{table}
The Wilson coefficients at the hadronic scale are computed within the SM at next-to-next-to-leading logarithms (NNLL) accuracy and are listed in Table~\ref{tab:SMWC1}~\cite{Altmannshofer19}. Furthermore, the Wilson coefficients in Eq.~\eqref{eq:WC12} are calculated at the matching scale $\mu_{\rm EW}$ and evolved down to the hadronic scale $\mu\sim m_{_b}$ by solving the renormalization group equations.
\begin{align}
\vec{C}_{NP}(\mu) &= \widehat{U}(\mu,\mu_{0})\, \vec{C}_{NP}(\mu_{0}) , \label{eq:C-evolution} \\
\vec{C}^{\prime}_{NP}(\mu) &= \widehat{U}^{\prime}(\mu,\mu_{0})\, \vec{C}^{\prime}_{NP}(\mu_{0}) .
\end{align}
with
\begin{align}
\overrightarrow{C}_{NP}^{T} &= \Bigl(
C_{1,NP},\;\cdots,\;C_{6,NP},
C_{7,NP}^{\,eff},\;C_{8,NP}^{\,eff},\;C_{9,NP}^{\,eff} - Y(q^{2}),\;
C_{10,NP}^{\,eff} \Bigr) , \nonumber \\
\overrightarrow{C}_{NP}^{\,\prime\,T} &= \Bigl(
C_{7,NP}^{\,\prime\,eff},\;
C_{8,NP}^{\,\prime\,eff},\;C_{9,NP}^{\,\prime\,eff},\;
C_{10,NP}^{\,\prime\,eff} \Bigr) .
\end{align}
Correspondingly, the evolving matrices are approached as
\begin{eqnarray}
	&&\widehat{U}(\mu,\mu_0)\simeq1-\Big[\frac{1}{2\beta_0}\ln\frac{\alpha_{_s}(\mu)}{\alpha_{_s}(\mu_0)}\Big]\widehat{\gamma}^{{\rm eff} (0),\;T}
	\;,\nonumber\\
	&&\widehat{U^\prime}(\mu,\mu_0)\simeq1-\Big[\frac{1}{2\beta_0}\ln\frac{\alpha_{_s}(\mu)}{\alpha_{_s}(\mu_0)}\Big]\widehat{\gamma^\prime}^{{\rm eff} (0),\;T}\;,
\end{eqnarray}
where the anomalous dimension matrices are calculated in Ref.~\cite{Gambino:NPB673}
\begin{eqnarray}
	&&\widehat{\gamma}^{{\rm eff} (0)}=\left(\begin{array}{cccccccccc}
		-4&\frac{8}{3}&0&-\frac{2}{9}&0&0&-\frac{208}{243}&\frac{173}{162}&-\frac{2272}{729}&0\\
		12&0&0&\frac{4}{3}&0&0&\frac{416}{81}&\frac{70}{27}&\frac{1952}{243}&0\\
		0&0&0&-\frac{52}{3}&0&2&-\frac{176}{81}&\frac{14}{27}&-\frac{6752}{243}&0\\
		0&0&-\frac{40}{9}&-\frac{100}{9}&\frac{4}{9}&\frac{5}{6}&-\frac{152}{243}&-\frac{587}{162}&-\frac{2192}{729}&0\\
		0&0&0&-\frac{256}{3}&0&20&-\frac{6272}{81}&\frac{6596}{27}&-\frac{84032}{243}&0\\
		0&0&-\frac{256}{9}&\frac{56}{9}&\frac{40}{9}&-\frac{2}{3}&\frac{4624}{243}&\frac{4772}{81}&-\frac{37856}{729}&0\\
		0&0&0&0&0&0&\frac{32}{3}&0&0&0\\
		0&0&0&0&0&0&-\frac{32}{9}&\frac{28}{3}&0&0\\
		0&0&0&0&0&0&0&0&0&0\\
		0&0&0&0&0&0&0&0&0&0\\
	\end{array}\right)
	\;,
\end{eqnarray}
\begin{eqnarray}
	&&\widehat{\gamma^\prime}^{{\rm eff} (0)}=\left(\begin{array}{cccc}
		\frac{32}{3}&0&0&0\\
		-\frac{32}{9}&\frac{28}{3}&0&0\\
		0&0&0&0\\0&0&0&0\\
	\end{array}\right)\;.
\end{eqnarray}

\section{Decay branching ratios and forward-backward asymmetries  \label{sec4}}
Using the above effective Hamiltonian and neglecting the s-quark mass, the matrix element for the quark level transition $b \to s l^{+} l^{-}$ is~\cite{Dai:PLB96}
\begin{equation}
\begin{aligned}[b]
\mathcal{M} &= \frac{\alpha G_F}{\sqrt{2}\pi} V_{tb} V_{ts}^* \Big\{
\bigl[ C_9 \bar{s}_L \gamma_\mu b_R + C_9' \bar{s}_R \gamma_\mu b_R \bigr] \bigl( \bar{l} \gamma^\mu l \bigr)
+ \bigl[ C_{10} \bar{s}_L \gamma_\mu b_R + C_{10}' \bar{s}_R \gamma_\mu b_L \bigr] \bigl( \bar{l} \gamma^\mu \gamma^5 l \bigr) \\
&\quad + 2i \frac{q^\nu}{q^2} \bigl[ \sigma_{\mu\nu} \bigl( C_7 \bar{s}_L m_b b_R + C_7' \bar{s}_R m_b b_L \bigr) \bigr] \bigl( \bar{l} \gamma^\mu l \bigr)
+ \bigl[ C_S \bar{s}_L b_R + C_S' \bar{s}_R b_L \bigr] \bigl( \bar{l} l \bigr) \\
&\quad + \bigl[ C_P \bar{s}_L b_R + C_P' \bar{s}_R b_L \bigr] \bigl( \bar{l} \gamma_5 l \bigr) \Big\}.
\end{aligned}
\label{M1}
\end{equation}
From the matrix element in Eq.~\eqref{M1}, the differential decay rate is derived as
\begin{equation}
\begin{aligned}[b]
\frac{d^{2}\Gamma}{dq^{2}d\cos\theta}
&= \frac{a_{\mathrm{EW}}^{2}}{16\pi^{2}}
\frac{G_{F}^{2}m_{b}^{5}|V_{tb}V_{ts}^{*}|^{2}}{48\pi^{3}}
\left(1-\frac{q^{2}}{m_{b}^{2}}\right)^{2}
\left(1-\frac{4m_{l}^{2}}{q^{2}}\right)^{\!1/2} \\
&\quad \times \Bigg\{
12\left(1+\frac{2m_{l}^{2}}{q^{2}}\right)
\Re\left(C_{7}^{\mathrm{eff}}C_{9}^{\mathrm{eff}*}(q^{2})\right)
+ \frac{6m_{l}^{2}}{m_{b}^{2}}
\left[|C_{9}^{\mathrm{eff}}(q^{2})|^{2} - |C_{10}^{\mathrm{eff}}|^{2}\right] \\
&\qquad + 6|C_{7}^{\mathrm{eff}}|^{2}
\left[1+\frac{m_{b}^{2}}{q^{2}}
- \left(1 -\frac{m_b^2}{q^2} -\frac{4m_l^2}{q^2}
+\frac{4m_l^2m_b^2}{q^4}\right)\cos^{2}\theta
+\frac{4m_l^2m_b^2}{q^4} \right] \\
&\qquad + \frac{3}{2}
\left[|C_{9}^{\mathrm{eff}}(q^{2})|^{2} + |C_{10}^{\mathrm{eff}}|^{2}\right]
\left[ 1+\frac{q^{2}}{m_{b}^{2}}
- \left(1 -\frac{q^2}{m_b^2} -\frac{4m_l^2}{q^2}
+\frac{4m_l^2}{m_b^2}\right)\cos^{2}\theta \right] \\
&\qquad - \frac{6m_{l}^{2}}{m_{b}^{2}} |C_{S}|^{2}
+ \frac{3q^{2}}{2m_{b}^{2}}
\left(|C_{S}|^{2}+|C_{P}|^{2}\right)\\
&\qquad + \left[\frac{6q^2}{m_{b}^2}
\Re\left(C_{9}^{\mathrm{eff}}(q^{2})C_{10}^{\mathrm{eff}*}\right)
+12\Re\left(C_{7}^{\mathrm{eff}}C_{10}^{\mathrm{eff}*}\right)\right]
\sqrt{1- \frac{4m_l^2}{q^{2}}} \cos\theta \\
&\qquad +\frac{6m_{l}}{m_{b}}
\Bigl[\Re\left(C_{9}^{\mathrm{eff}}(q^{2})C_{S}^{*}
+2C_{7}^{\mathrm{eff}}C_{S}^{*}\right)
\sqrt{1- \frac{4m_l^2}{q^{2}}}\cos\theta
+ \Re\left(C_{10}C_{P}^{*}\right) \Bigr] \Bigg\}.
\end{aligned}
\label{T1}
\end{equation}
The normalized differential branching ratio for $B \to X_{\mathrm{s}} l^{+} l^{-}$ is obtained as
\begin{equation}
\begin{aligned}[b]
R(q^{2}) = \frac{d\mathrm{Br}(B \to X_{\mathrm{s}} l^{+} l^{-})}{\Gamma(B\rightarrow X_{c}e\nu)dq^{2}}
&=\frac{a_{\mathrm{EW}}^{2}}{2\pi^{2}} \frac{|V_{tb} V_{ts}^{*}|^{2}}{|V_{cb}|^{2}}
\frac{1}{f(z)\kappa(z)}
\left(1 - \frac{q^{2}}{m_{b}^{2}}\right)^{2} \sqrt{1 - \frac{4m_{l}^{2}}{q^{2}}} \\
&\quad \times \Bigg\{
12\left(1 + \frac{2m_{l}^{2}}{q^{2}}\right) \Re(C_{7}^{\mathrm{eff}} C_{9}^{\mathrm{eff}*}(q^{2})) \\
&\quad + 4|C_{7}^{\mathrm{eff}}|^{2} \left[1 + 2\frac{m_{b}^{2}}{q^{2}} + \frac{2m_{l}^{2}}{q^{2}} + \frac{4 m_{l}^{2} m_{b}^{2}}{q^{4}}\right] \\
&\quad + |C_{9}^{\mathrm{eff}}(q^{2})|^{2} \left[1 + 2\frac{q^{2}}{m_{b}^{2}} + \frac{2m_{l}^{2}}{q^{2}} + \frac{4m_{l}^{2}}{m_{b}^{2}}\right] \\
&\quad + |C_{10}^{\mathrm{eff}}|^{2} \left[1 + 2\frac{q^{2}}{m_{b}^{2}} + \frac{2m_{l}^{2}}{q^{2}} - \frac{8m_{l}^{2}}{m_{b}^{2}}\right] \\
&\quad -\frac{6m_{l}^{2}}{m_{b}^{2}}|C_{S}|^{2} + \frac{3q^{2}}{2m_{b}^{2}}\big(|C_{S}|^{2}+|C_{P}|^{2}\big) \\
&\quad + \frac{6m_{l}}{m_{b}}\Re(C_{10}^{\mathrm{eff}} C_{P}^{*}) \Bigg\}.
\label{RQ}
\end{aligned}
\end{equation}
The normalized differential branching ratio is defined using the $B$-meson semileptonic decay width, which reduces uncertainties from the bottom quark mass and CKM matrix elements. The corresponding formula is given below~\cite{Ghincuulov:NPB685}
\begin{equation}
\Gamma(B\to X_{c}e\bar{\nu}_{e})=\frac{G_{F}^{2}m_{b}^{5}}{192\pi^{3}}|V_{cb}|^{2}f(z)\kappa(z).
\label{eq:Gamma_BXc11}
\end{equation}
Here, $f(z)=1-8z^{2}+8z^{6}-z^{8}-24z^{4}\ln z$ and $\kappa(z)=1-\frac{2\alpha_{s}}{3\pi}\bigl[(\pi^{2}-\tfrac{31}{4})(1-z)^{2}+\tfrac{3}{2}\bigr]$ are the phase space factor and the QCD correction factor ($z=m_{c}/m_{b}$)~\cite{Buchalla:RMP68}.
The unnormalized forward-backward asymmetry is expressed as
\begin{align}
\bar{A}_{FB}(q^{2}) & = \frac{1}{\Gamma(B\rightarrow X_{c}e\nu)}
\int_{-1}^{1} d\cos\theta\,
\frac{d^{2}\Gamma(B \to X_{\mathrm{s}} l^{+} l^{-})}{d\cos\theta\,dq^{2}}
\operatorname{sgn}(\cos\theta) \nonumber \\
& = \frac{3\alpha_{\mathrm{EW}}^{2}}{2\pi^{2}}
\biggl|\frac{V_{tb}V_{ts}^{*}}{V_{cb}}\biggr|^{2}
\frac{1}{f(z)\kappa(z)}
\Bigl(1-\frac{q^{2}}{m_{b}^{2}}\Bigr)^{\!2}
\Bigl(1-\frac{4m_{l}^{2}}{q^{2}}\Bigr) \nonumber \\
& \quad \times \Biggl\{
\frac{q^{2}}{m_{b}^{2}}\,\Re\bigl(C_{9}^{\mathrm{eff}}(q^{2})C_{10}^{\mathrm{eff}*}\bigr)
+ 2\Re\bigl(C_{7}^{\mathrm{eff}}C_{10}^{\mathrm{eff}*}\bigr) \nonumber \\
& \quad\qquad
+ \frac{m_{l}}{m_{b}}\,\Re\bigl(C_{9}^{\mathrm{eff}}(q^{2})C_{S}^{*}
+ 2C_{7}^{\mathrm{eff}}C_{S}^{*}\bigr)
\Biggr\},
\label{eq:Unnormalized-AFB}
\end{align}
and the normalized forward-backward asymmetry is written as
\begin{equation}
A_{FB}(q^{2}) = \frac{1}{R(q^{2})} \bar{A}_{FB}(q^{2}) ,
\label{eq:normalized-AFB}
\end{equation}
The global forward-backward asymmetry in the region $q^2\in[a,\;b]\;{\rm GeV}^2$
is given by~\cite{Scimemi:NPB81}
\begin{align}
\frac{N(\ell_{+}^{+}) - N(\ell_{+}^{-})}{N(\ell_{+}^{+}) + N(\ell_{+}^{-})}
& \equiv
\frac{
    \displaystyle\int_{-1}^{1} \mathrm{d}(\cos\theta) \int \mathrm{d}s\,
    \frac{\mathrm{d}^{2} \Gamma(B \to X_{s} \ell^{+}\ell^{-})}
         {\mathrm{d}(\cos\theta) \,\mathrm{d}s}
    \operatorname{sgn}(\cos\theta)
}{
    \displaystyle\int_{-1}^{1} \mathrm{d}(\cos\theta) \int \mathrm{d}s\,
    \frac{\mathrm{d}^{2} \Gamma(B \to X_{s} \ell^{+}\ell^{-})}
         {\mathrm{d}(\cos\theta) \,\mathrm{d}s}
}.
\end{align}
While these definitions apply generally over the full dilepton invariant-mass spectrum, the theoretical treatment differs significantly between the low- and high-$q^2$ regions. Theoretical predictions for observables in $B \to X_{\mathrm{s}} l^{+} l^{-}$ decays at large $q^2$ receive significant nonperturbative corrections. In particular, the $1/m_b^3$ contributions are not parametrically suppressed relative to the $1/m_b^2$ terms~\cite{Ligeti:PLB653}. In the high-$q^2$ region, the operator product expansion (OPE) effectively becomes an expansion in powers of $\Lambda_{\mathrm{QCD}}/(m_b-\sqrt{q^2})$ rather than $\Lambda_{\mathrm{QCD}}/m_b$, thereby enhancing hadronic uncertainties. As noted in Ref.~\cite{Huber:NPB802}, the heavy-quark expansion breaks down near the kinematic endpoint of the inclusive decay. In this region, no shape-function resummation is possible, limiting the reliability of perturbative predictions. Physical observables, such as the branching ratio and the forward-backward asymmetry, involve integrals over $|\Pi(q^2)|^2$ or related combinations. Moreover, global quark-hadron duality is not expected to hold for $|\Pi(q^2)|^2$ in this region~\cite{Beneke:EPJC61}, introducing additional hadronic uncertainties. Recent analyses~\cite{Isidori:PRD108} indicate that even within a semi-inclusive framework, high-$q^2$ observables remain affected by residual nonlocal charm effects and form-factor contributions. Although these effects are typically at the $\mathcal{O}(1\%)$ level, they limit the precision of conventional perturbative predictions for the decay rate. In this situation, an analysis based on angular coefficients provides a more transparent theoretical description of the forward-backward asymmetry, since these observables admit a direct interpretation in terms of transversity amplitudes.

Accordingly, we employ the angular coefficient formalism for $B\to K^*\mu^+\mu^-$ decays developed in Ref.~\cite{Altmannshofer19}, while adopting the opposite sign convention for $I_6^s$ to match the high-$q^2$ kinematics. With this convention, we reproduce the Standard Model predictions shown in Table III. Within this framework, the forward-backward asymmetry is expressed in terms of the CP-averaged angular coefficients $S_6^s$ and $S_6^c$ as
\begin{equation}
A_{\rm FB} = \frac{3}{8} (2 S_6^{s} + S_6^{c}) \,,
\end{equation}
(see Eqs.~(5.7) of Ref.~\cite{Altmannshofer19}). These coefficients are constructed from transversity amplitudes, which factorize into short-distance Wilson coefficients and form factors in the heavy-quark limit, thereby allowing for a more controlled treatment of hadronic uncertainties in the high-$q^2$ region.

The SM predictions for $\mathrm{BR}(B \to X_{\mathrm{s}} e^{+} e^{-})$ and $\mathrm{BR}(B \to X_{\mathrm{s}} \mu^{+} \mu^{-})$ in the low-$q^2$ region, including nonperturbative effects through an updated Kr\"uger-Sehgal (KS) analysis, are~\cite{Huber:JHEP10SMP}
\begin{align}
&\mathrm{BR}(B \to X_{\mathrm{s}} e^{+} e^{-})_{q^2 \in [1,6]\,\mathrm{GeV}^2}^{\mathrm{SM}}
=(1.78\pm0.13)\times10^{-6}\;,\nonumber\\
&\mathrm{BR}(B \to X_{\mathrm{s}} \mu^{+} \mu^{-})_{q^2 \in [1,6]\,\mathrm{GeV}^2}^{\mathrm{SM}}
=(1.73\pm0.13)\times10^{-6}\;.
\label{BR1}
\end{align}
The SM predictions for $\mathrm{BR}(B \to X_{\mathrm{s}} e^{+} e^{-})$ and $\mathrm{BR}(B \to X_{\mathrm{s}} \mu^{+} \mu^{-})$ in the high-$q^2$ region are given by~\cite{Huber:JHEP10SMP}
\begin{align}
&\mathrm{BR}(B \to X_{\mathrm{s}} e^{+} e^{-})_{q^2 \in [14.4,25]\,\mathrm{GeV}^2}^{\mathrm{SM}}
=(2.04\pm0.87)\times10^{-7}\;,\nonumber\\
&\mathrm{BR}(B \to X_{\mathrm{s}} \mu^{+} \mu^{-})_{q^2 \in [14.4,25]\,\mathrm{GeV}^2}^{\mathrm{SM}}
=(2.38\pm0.87)\times10^{-7}\;.
\label{BR2}
\end{align}
In our analysis, the lepton-flavor-averaged branching ratio for $B \to X_{\mathrm{s}} l^{+} l^{-}$ is obtained by averaging the branching ratios of the electron and muon final states.

\section{NUMERICAL ANALYSIS\label{sec5}}
In this section, we will present the numerical results of the rare $B$ meson decay $B \to X_{\mathrm{s}} l^{+} l^{-}$ and analyze its underlying physical mechanisms. To obtain clear numerical results, we adopt specific assumptions~\cite{Ambrosio:NPB645} for certain parameters in the $\mu\nu$SSM. These assumptions are as follows:
\begin{align}
&\hspace{-0.9cm}\kappa_{ijk} = \kappa \delta_{ij} \delta_{jk}, \quad
(A_\kappa \kappa)_{ijk} = A_\kappa \kappa \delta_{ij} \delta_{jk}, \quad
\lambda_i = \lambda, \nonumber\\
&\hspace{-0.9cm}(A_\lambda \lambda)_i = A_\lambda \lambda, \quad
Y_{e_{ij}} = Y_{e_i} \delta_{ij}, \quad
(A_e Y_e)_{ij} = A_e Y_{e_i} \delta_{ij}, \nonumber\\
&\hspace{-0.9cm}Y_{\nu_{ij}} = Y_{\nu_i} \delta_{ij}, \quad
(A_\nu Y_\nu)_{ij} = a_{\nu_i} \delta_{ij}, \quad
m_{\tilde{\nu}_{ij}^c}^2 = m_{\tilde{\nu}_i^c}^2 \delta_{ij}, \nonumber\\
&\hspace{-0.9cm}m_{\tilde{L}_{ij}}^2 = m_{\tilde{L}}^2 \delta_{ij}, \quad
m_{\tilde{e}_{ij}^c}^2 = m_{\tilde{e}^c}^2 \delta_{ij}, \quad
\upsilon_{\nu_i^c} = \upsilon_{\nu^c}.
\end{align}
where $i,j,k=1,2,3$. Constrained by the masses of quarks and leptons, one can define the following expression:
\begin{equation}
Y_{e}^{i} = \frac{m_{l}^{i}}{\upsilon_{d}},\quad
Y_{u}^{i} = \frac{m_{u}^{i}}{\upsilon_{u}},\quad
Y_{d}^{i} = \frac{m_{d}^{i}}{\upsilon_{d}}.
\end{equation}
where $m_{l}^{i}$, $m_u^i$ and $m_d^i$ represent the masses of charged leptons, up-quarks and down-quarks, respectively. The specific values can be obtained from PDG~\cite{PDG:2024PDG13}. Additionally, the soft masses $m_{H_d}^2$, $m_{H_u}^2$, $m_{\tilde\nu^c_{ij}}^2$, and $m_{\tilde{L}_{ij}}^2$ can be calculated from the minimization condition of tree-level neutral scalar potential, and their explicit forms are provided in Ref.~\cite{Escudero:JHEP08}.

In our earlier work~\cite{neutrino-mass13}, the Yukawa coupling $Y_{\nu_{i}} \sim \mathcal{O}(10^{-7})$ and left-handed sneutrino VEVs $\upsilon_{\nu_{i}} \sim \mathcal{O}(10^{-4}\,\text{GeV})$ are governed by the TeV seesaw mechanism. The corresponding contributions to the theoretical predictions for $B \to X_{\mathrm{s}} l^{+} l^{-}$ observables in both the low- and high-$q^2$ regions are typically at the level of $\mathcal{O}(10^{-5})$.   Therefore, these effects are neglected in the present numerical analysis, and we can approximately set $Y_{\nu_{i}}=0$, and $\upsilon_{\nu_{i}}=0$. By analyzing the parameter space of the $\mu\nu$SSM in Ref.~\cite{Escudero:JHEP08}, and for simplicity,  we choose the suitable parameter values $\lambda = 0.135$, $A_{\kappa} = -300\,\mathrm{GeV}$, $m_{\tilde{L}_{1,2,3}} = m_{\tilde{e}_{1,2,3}^{c}} = 700\,\mathrm{GeV}$ and $A_{u_{1,2}} = A_{d_{1,2,3}} = A_{e_{1,2,3}} = 1\,\mathrm{TeV}$. Considering the strong constraints from direct searches at the LHC~\cite{ATLAS:PRD87,CMS.JHEP1210}, we take $m_{\tilde{Q}_{1,2,3}} = m_{\tilde{u}_{1,2}^{c}} = m_{\tilde{d}_{1,2,3}^{c}} = 3\,\mathrm{TeV}$. Following the approximate GUT relation, we set the gaugino Majorana masses as $M_{1} = 0.5 M_{2}$ and $M_{3} = 2.7 M_{2}$. In R-parity violating SUSY models, the gluino mass $m_{\tilde g} \approx M_3$ is constrained to be $> 2.4\,\mathrm{TeV}$~\cite{PDG:2024PDG13,gluinoATLAS21}. Considering the above constraints, we choose $M_{2} = 1\,\mathrm{TeV}$,  and the gluino mass $m_{\tilde{g}} = 2.7\,\mathrm{TeV}$. Given the significant influence of $A_{t}$ and $m_{\tilde{u}_{3}^{c}}$ on the SM Higgs mass, we choose $A_{t} = 2.6\,\mathrm{TeV}$, and $m_{\tilde{u}_{3}^{c}} = 3\,\mathrm{TeV}$ to obtain $m_{h} \approx 125\,\mathrm{GeV}$. The relevant SM input parameters are summarized in Table~\ref{tab:dominant parameters11}.

Moreover, the measured average branching ratios for $B_{\mathrm{s}}^{0} \to \mu^{+} \mu^{-}$ and $\bar{B} \to X_{\mathrm{s}}\gamma$ are expressed as~\cite{PDG:2024PDG13,HFLAV:PRD107,ATLAS:JHEP04,CMS:JHEPbbr,LHCb:PRL128,CMS:PLB842}
\begin{align}
	&\mathrm{BR}(B_{\mathrm{s}}^{0} \to \mu^{+} \mu^{-}) = (3.34 \pm 0.27) \times 10^{-9}, \nonumber \\
&\mathrm{BR}(\bar{B} \to X_{\mathrm{s}}\gamma) = (3.49 \pm 0.19) \times 10^{-4}.
     \label{E1rl}
\end{align}
The SM predicts that the branching ratios for $B_{\mathrm{s}}^{0} \to \mu^{+} \mu^{-}$ and $\bar{B} \to X_{\mathrm{s}}\gamma$  are~\cite{Misiak:JHEP06,Beneke:PRL120,Beneke:JHEP10,Czaja:SY16}
\begin{align}
	&\mathrm{BR}(B_{\mathrm{s}}^{0} \to \mu^{+}\mu^{-}) = (3.64 \pm 0.12) \times 10^{-9}, \nonumber \\
&\mathrm{BR}(\bar{B} \to X_{\mathrm{s}}\gamma) = (3.40 \pm 0.17) \times 10^{-4}.
\label{SM2rl}
\end{align}
\begin{table}[htbp]
\centering
\caption{Input parameters of the SM ~\cite{PDG:2024PDG13} used in the numerical analysis.}
\begin{tabular}{ll}
\hline\hline
input &input \\
\hline
\( m_B = 5279.63 \pm 0.20 \, \text{MeV} \) & \( m_{K^*} = 0.896 \, \text{GeV} \) \\
\( m_{B_s^0} = 5366.93 \pm 0.10 \, \text{MeV} \) & \( m_{\mu} = 105.6583755 \pm 0.0000023 \, \text{MeV} \) \\
\( m_W = 80.3692 \pm 0.0133 \, \text{GeV} \) & \( m_z = 91.1880 \pm 0.0020  \, \text{GeV} \) \\
\(\tau_B = (1.519 \pm 0.004)\,\mathrm{ps}\) \cite{HFLAV:PRD107} & \( \alpha_s(m_Z) = 0.1180(9) \)\\
\(\tau_{B_s^0} = (1.527 \pm 0.011)\,\mathrm{ps}\) \cite{HFLAV:PRD107} & \( \alpha_{\text{em}}(m_Z) = 1/127.944(14) \) \\
\( m_c(m_c) = 1.2730 \pm 0.0046 \, \text{GeV} \) & \( m_b(m_b) = 4.183 \pm 0.007 \, \text{GeV} \) \\
\( m_t^{\text{pole}} = 172.4 \pm 0.7 \, \text{GeV} \) & \\
\hline
\end{tabular}
\label{tab:dominant parameters11}
\end{table}

To validate the numerical framework, all new physics contributions are systematically switched off to recover the Standard Model (SM) limit. Using the input parameters listed in Table~\ref{tab:SMWC1}, we evaluate the branching ratios $\mathrm{BR}(B \to X_{\mathrm{s}} l^{+} l^{-})$ and the forward-backward asymmetry $A_{\mathrm{FB}}(B \to X_{\mathrm{s}} l^{+} l^{-})$ for the $b\to s\ell^+\ell^-$ transition in low- and high-$q^2$ regions. The branching ratios $\mathrm{BR}(B \to X_{\mathrm{s}} l^{+} l^{-})$ are computed using the standard approximate parametrization of Ref.~\cite{Hurth:NPB808}. The corresponding numerical results are summarized in Table~\ref{tab:SMbenchmark}. The obtained predictions are found to be consistent within the $1\sigma$ uncertainty range with the SM results discussed in the Introduction, thereby validating the numerical framework adopted in the present work.
\begin{table}[htbp]
\centering
\caption{SM predictions for $B \to X_{\mathrm{s}} l^{+} l^{-}$ observables.}
\begin{tabular}{lcc}
\hline
Observable & Our SM prediction \\
\hline
$\mathrm{BR}(B \to X_{\mathrm{s}} l^{+} l^{-})_{q^2 \in [1,6]\, \mathrm{GeV}^2}$ &  $1.6\times 10^{-6}$\\
$\mathrm{BR}(B \to X_{\mathrm{s}} l^{+} l^{-})_{q^2 \in [14.4,25]\, \mathrm{GeV}^2}$ &  $3.0\times 10^{-7}$\\
$\mathrm{A}_{\rm FB}(B \to X_{\mathrm{s}} l^{+} l^{-})_{q^2 \in [1,6]\, \mathrm{GeV}^2}$& $-0.097$ \\
$\mathrm{A}_{\rm FB}(B \to X_{\mathrm{s}} l^{+} l^{-})_{q^2 \in [14.4,25]\, \mathrm{GeV}^2}$ & 0.377 \\
\hline
\end{tabular}
\label{tab:SMbenchmark}
\end{table}

To avoid tachyons in the sneutrino sector, an analysis of the sneutrino mass spectrum is required. Specifically, the left-handed sneutrinos masses are predominantly determined by soft SUSY-breaking mass $m_{\tilde{L}}$, whereas the three right-handed sneutrinos remain essentially degenerate. The squared masses of CP-even and CP-odd right-handed sneutrinos can be approximated as~\cite{Zhang:PRD14}:
 \begin{align}
m_{S_{5+i}}^{2} &\approx (A_{\kappa} + 4\kappa \upsilon_{\nu^{c}})\kappa \upsilon_{\nu^{c}}
                 + A_{\lambda} \lambda \upsilon_{d} \upsilon_{u}/\upsilon_{\nu^{c}}
                 - 2\lambda^{2}(\upsilon_{d}^{2} + \upsilon_{u}^{2}), \label{eqMmass1} \\
m_{P_{5+i}}^{2} &\approx -3 A_{\kappa} \kappa \upsilon_{\nu^{c}}
                 + (A_{\lambda}/\upsilon_{\nu^{c}} + 4\kappa)\lambda \upsilon_{d} \upsilon_{u}
                 - 2\lambda^{2}(\upsilon_{d}^{2} + \upsilon_{u}^{2}). \label{eqMmass2}
\end{align}
Eqs.~\eqref{eqMmass1} and ~\eqref{eqMmass2} indicate that for large $\kappa$ and in the limit $\upsilon_{\nu^{c}} \gg \upsilon_{u,d}$, the dominant contribution to the mass squared of the right-handed sneutrinos arises from the first term. Therefore, we can adopt the approximate relation
\begin{align}
-4\kappa \upsilon_{\nu^{c}} \lesssim A_{\kappa} \lesssim 0,
\label{tachyon}
\end{align}
to avoid the emergence of tachyons. While $A_{\kappa}$ has minimal influence on numerical calculations, the subsequent numerical analysis shows that the parameters $\kappa$ and $\upsilon_{\nu^c}$ have a visible effect on the calculations.

In the $\mu\nu$SSM, the charged Higgs mass squared, $M_{H^{\pm}}^{2}$, in the limit of $\upsilon_{\nu^{c}} \gg \upsilon_{u,d}$ can be written as
\begin{align}
M_{H^{\pm}}^{2} & \simeq M_{A}^{2} + \Bigl( 1 - \frac{6 s_{W}^{2} \lambda^{2}}{e^{2}} \Bigr) m_{W}^{2},
\label{eq:MHc1}
\end{align}
where the squared mass of the neutral pseudoscalar is
\begin{align}
M_{A}^{2} & \simeq \frac{6 \lambda \upsilon_{\nu^{c}} (A_{\lambda} + \kappa \upsilon_{\nu^{c}})}{\sin 2\beta}.
\label{eq:MA2}
\end{align}

\begin{figure}[t]
  \centering
  \includegraphics[width=\columnwidth]{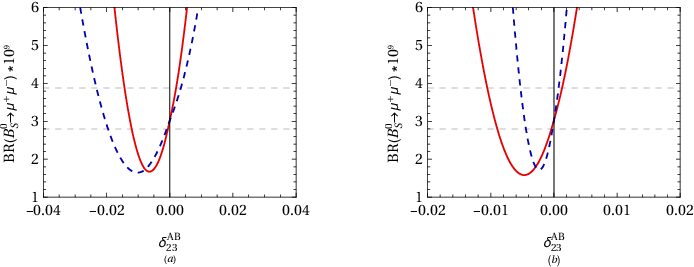}
  \caption{
Branching ratio $\mathrm{BR}(B_{\mathrm{s}}^{0} \to \mu^{+} \mu^{-})$
as a function of the mass insertions
$\delta_{23}^{AB}$,with (a) $\tan\beta = 20$ (solid line) and $\tan\beta = 30$ (dashed line), and (b)$\kappa = 0.06$ (solid line) and $\kappa = 0.3$ (dashed line).
}
  \label{fig2:sixplots}
\end{figure}

Before presenting the numerical results, we briefly comment on the experimental constraints relevant for our parameter scan.  Experimental constraints from the decay process of $B \to X_{\mathrm{s}} l^{+} l^{-}$, $\bar{B} \to X_{\mathrm{s}}\gamma$ and $B_{\mathrm{s}}^{0} \to \mu^{+} \mu^{-}$ impose stringent restrictions on the allowed parameter space. Among these, the purely leptonic decay $B_{\mathrm{s}}^{0} \to \mu^{+} \mu^{-}$ provides the most stringent constraints on the flavor-violating parameters $\delta_{23}^{AB}$. To determine the viable values of these parameters (with $\tan\beta$ and $\kappa$ as auxiliary parameters), we perform a systematic scan over the parameter space. The branching ratio $\mathrm{BR}(B_{\mathrm{s}}^{0} \to \mu^{+} \mu^{-})$ as a function of the mass insertion $\delta_{23}^{AB}$ is shown in Fig. \ref{fig2:sixplots}, with the  2$\sigma$ experimental bounds overlaid. In Fig. \ref{fig2:sixplots}(a), results are shown for $\kappa = 0.4$, $A_\lambda = 0.6~\mathrm{TeV}$, and $\upsilon_{\nu^c} = 2~\mathrm{TeV}$ with $\tan\beta = 20$ (solid line) and $\tan\beta = 30$ (dashed line). Fig. \ref{fig2:sixplots}(b) shows results for $\tan\beta = 25$, $A_\lambda = 0.5~\mathrm{TeV}$, and $\upsilon_{\nu^c} = 2~\mathrm{TeV}$ with $\kappa = 0.06$ (solid line) and $\kappa = 0.3$ (dashed line).

\begin{figure}[b]
  \centering
  \includegraphics[width=\columnwidth]{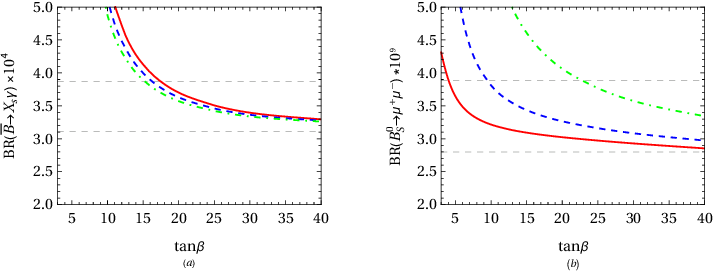}
   \caption{ Fixed $\kappa=0.4$, $A_\lambda = 0.6~\mathrm{TeV}$: (a) $\mathrm{BR}(\bar{B} \to X_{\mathrm{s}}\gamma)$ and (b) $\mathrm{BR}(B_{\mathrm{s}}^{0} \to \mu^{+} \mu^{-})$  vary with $\tan\beta$ for $\upsilon_{\nu^c}=1.8~\mathrm{TeV}$ (solid line), $\upsilon_{\nu^c}=2.0~\mathrm{TeV}$ (dashed line), and $\upsilon_{\nu^c}=2.2~\mathrm{TeV}$ (dashed-dotted line). }
  \label{fig3:sixplots}
\end{figure}

\begin{table}[htbp]
\setlength{\tabcolsep}{3pt}
\begin{tabular}{l l l l}
\hline\hline
Observable & leading Particle & leading WCS & leading Diagram \\
\hline
$\mathrm{BR}(\bar{B} \to X_{\mathrm{s}}\gamma)$ & $H^\pm$ & $C_7$ & $\gamma$-penguin \\
$\mathrm{BR}(B_{\mathrm{s}}^{0} \to \mu^{+} \mu^{-})$ & $\tilde\chi^\pm$ & $C_S$ & $S$-penguin \\
$\mathrm{BR}(B \to X_{\mathrm{s}} l^{+} l^{-})_{q^2 \in [1,6]\, \mathrm{GeV}^2}$ & $H^\pm$ & $C_7$ & $\gamma$-penguin \\
$\mathrm{BR}(B \to X_{\mathrm{s}} l^{+} l^{-})_{q^2 \in [14.4,25]\, \mathrm{GeV}^2}$ & $H^\pm,\tilde\chi^\pm$ & $C_9, C_{10}$ & Z-penguin \\
$\mathrm{\bar{A}}_{\rm FB}(B \to X_{\mathrm{s}} l^{+} l^{-})_{q^2 \in [1,6]\, \mathrm{GeV}^2}$ & $H^\pm$& $C_7C_{10}$, $C_9C_{10}$ & $\gamma$-penguin and $Z$-penguin \\
$\mathrm{A}_{\rm FB}(B \to X_{\mathrm{s}} l^{+} l^{-})_{q^2 \in [14.4,25]\, \mathrm{GeV}^2}$& $H^\pm$ & $C_7C_{10}$, $C_9C_{10}$ &$\gamma$-penguin and $Z$-penguin\\
\hline\hline
\end{tabular}
\caption{Summary of dominant contributions to $b\to s\ell^+\ell^-$ observables in two different $q^2$ regions.}
\label{tab:dominant_contributions}
\end{table}
In this analysis, we investigate the rare decay $B \to X_{\mathrm{s}} l^{+} l^{-}$ by scanning the relevant parameters and studying their impact on the observables. Since the flavor-violating parameter $\delta_{23}^{AB}$ is constrained by the experimental measurement of $B_s^0 \to \mu^+ \mu^-$, we adopt $\delta_{23}^{AB} = 0.6 \times 10^{-3}$ as the benchmark value when scanning $\tan\beta$, with an allowed variation in the range $(0.6 - 1.4) \times 10^{-3}$. For the $\kappa$ scan, we take $\delta_{23}^{AB} = 0.20 \times 10^{-3}$ as the benchmark value, varying within approximately $(0.14 - 0.22) \times 10^{-3}$.

The parameter $\tan\beta$ is also mainly constrained by the $\bar{B} \to X_{\mathrm{s}}\gamma$  decay. Fig. \ref{fig3:sixplots} shows the allowed region for $\tan\beta$ at the $2\sigma$ confidence level, subject to the constraint on the SM-like Higgs mass $(124\text{--}126\,\mathrm{GeV})$. In this allowed region, $B \to X_{\mathrm{s}} l^{+} l^{-}$ observables are  within the $3\sigma$ experimental ranges and therefore do not provide additional constraints. Specifically, $\kappa$ is scanned from $0.01$ to $0.6$. The Higgs mass requirement restricts $\kappa$ to the narrow range of $0.4-0.6$, and no other observables further constrain this interval. Values of $\kappa$ are included only to demonstrate the behavior of observables outside the viable region.

To analyze the underlying physics of $B \to X_{\mathrm{s}} l^{+} l^{-}$ decay, we calculate the relevant Feynman diagrams (representative diagrams are shown in Fig.~\ref{fig:feynman}) and determine the corresponding contributions of new physics to the Wilson coefficients. Theoretical predictions for these coefficients can then be confronted with experimental data on $\mathrm{BR}(B \to X_{\mathrm{s}} l^{+} l^{-})$ and $A_{\mathrm{FB}}(B \to X_{\mathrm{s}} l^{+} l^{-})$, allowing us to identify which coefficients have the most significant impact. The dominant new physics contributions to the $B$-meson observables are identified and summarized in Table~\ref{tab:dominant_contributions}.
\begin{figure}[t]
  \centering
  \includegraphics[width=\columnwidth]{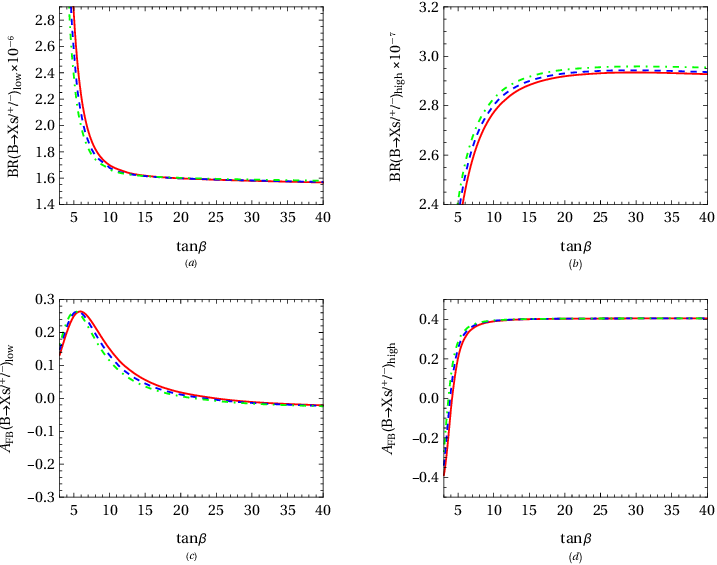}
   \caption{
(a) $\mathrm{BR}(B \to X_{\mathrm{s}} l^{+} l^{-})_{q^2 \in [1,6]~\mathrm{GeV}^2}$,
(b) $\mathrm{BR}(B \to X_{\mathrm{s}} l^{+} l^{-})_{q^2 \in [14.4,25]~\mathrm{GeV}^2}$,
(c) $A_{\mathrm{FB}}(B \to X_{\mathrm{s}} l^{+} l^{-})_{q^2 \in [1,6]~\mathrm{GeV}^2}$, and
(d) $A_{\mathrm{FB}}(B \to X_{\mathrm{s}} l^{+} l^{-})_{q^2 \in [14.4,25]~\mathrm{GeV}^2}$ as functions of $\tan\beta$ for fixed $\kappa=0.4$, $A_\lambda = 0.6~\mathrm{TeV}$, with $\upsilon_{\nu^c}=1.8~\mathrm{TeV}$ (solid line),
$\upsilon_{\nu^c}=2.0~\mathrm{TeV}$ (dashed line), and $\upsilon_{\nu^c}=2.2~\mathrm{TeV}$ (dashed-dotted line).
}
\label{fig4:sixplots}
\end{figure}
Using the chosen parameter space, we present $\mathrm{BR}(B \to X_{\mathrm{s}} l^{+} l^{-})$ in the low-$q^2$ (1-6 $\mathrm{GeV}^2$) and high-$q^2$ (14.4-25 $\mathrm{GeV}^2$) regions as functions of $\tan\beta$ in Fig. \ref{fig4:sixplots}(a) and Fig. \ref{fig4:sixplots}(b), respectively. The corresponding $A_{\mathrm{FB}}(B \to X_{\mathrm{s}} l^{+} l^{-})$ in the low- and high-$q^2$ regions are shown in Fig. \ref{fig4:sixplots}(c) and Fig. \ref{fig4:sixplots}(d), respectively. For brevity, we define $\mathrm{BR}(B \to X_{\mathrm{s}} l^{+} l^{-})$ in two different $q^2$ regions as
\begin{align}
\mathrm{BR}(B \to X_{\mathrm{s}} l^{+} l^{-})_{\mathrm{low}} &= \mathrm{BR}(B \to X_{\mathrm{s}} l^{+} l^{-})_{q^2 \in [1,6]\, \mathrm{GeV}^2}, \\
\mathrm{BR}(B \to X_{\mathrm{s}} l^{+} l^{-})_{\mathrm{high}} &= \mathrm{BR}(B \to X_{\mathrm{s}} l^{+} l^{-})_{q^2 \in [14.4,25]\, \mathrm{GeV}^2}.
\end{align}
It is evident that the branching ratios of $B \to X_{\mathrm{s}} l^{+} l^{-}$ exhibit distinct trends in the low- and high-$q^2$ regions. From Fig. \ref{fig4:sixplots}(a), we observe that $\mathrm{BR}(B \to X_{\mathrm{s}} l^{+} l^{-})_{\mathrm{low}}$ decreases gradually with increasing $\tan\beta$, while exhibiting only a mild dependence on the parameter $\upsilon_{\nu^c}$. The theoretical prediction for $\mathrm{BR}(B \to X_{\mathrm{s}} l^{+} l^{-})_{\mathrm{low}}$ decreases from $ 2.8 \times 10^{-6} $ to $ 1.6 \times 10^{-6} $ as $\tan\beta$ increases. In contrast, Fig. \ref{fig4:sixplots}(b) shows that $\mathrm{BR}(B \to X_{\mathrm{s}} l^{+} l^{-})_{\mathrm{high}}$ exhibits a different behavior as $\tan\beta$ increases, increasing from $ 2.3 \times 10^{-6} $ to $ 2.9 \times 10^{-6}$. The differing trends observed in these two $q^2$ intervals imply that the corresponding contributions may originate from distinct sources.

By comparison, the forward-backward asymmetries exhibit a different pattern. Similarly, we define  $A_{\mathrm{FB}}(B \to X_{\mathrm{s}} l^{+} l^{-})$ in the two $q^2$ regions as
\begin{align}
A_{\mathrm{FB}}(B \to X_{\mathrm{s}} l^{+} l^{-})_{\mathrm{low}} &= A_{\mathrm{FB}}(B \to X_{\mathrm{s}} l^{+} l^{-})_{q^2 \in [1,6]\,\mathrm{GeV}^2}, \\
A_{\mathrm{FB}}(B \to X_{\mathrm{s}} l^{+} l^{-})_{\mathrm{high}} &= A_{\mathrm{FB}}(B \to X_{\mathrm{s}} l^{+} l^{-})_{q^2 \in [14.4,25]\,\mathrm{GeV}^2}.
\end{align}
From Fig. \ref{fig4:sixplots}(c) and Fig. \ref{fig4:sixplots}(d), we observe that $A_{\mathrm{FB}}(B \to X_{\mathrm{s}} l^{+} l^{-})_{\mathrm{low}}$ first increases and then decreases as $\tan\beta$ grows, while in the high-$q^2$ region $A_{\mathrm{FB}}(B \to X_{\mathrm{s}} l^{+} l^{-})_{\mathrm{high}}$ exhibits an overall increasing trend. Specifically, as $\tan\beta$ increases, $A_{\mathrm{FB}}(B \to X_{\mathrm{s}} l^{+} l^{-})_{\mathrm{low}}$ varies within the range from $-0.1$ to $0.2$, while $A_{\mathrm{FB}}(B \to X_{\mathrm{s}} l^{+} l^{-})_{\mathrm{high}}$ increases from $-0.4$ to $0.4$. In addition, in the low-$q^2$ region the theoretical predictions for $A_{\mathrm{FB}}(B \to X_{\mathrm{s}} l^{+} l^{-})$  are not significantly affected by the photon pole at low-$q^2$ or by the $c\bar{c}$ resonance contributions at higher $q^2$. The distinct evolutionary behaviors observed in the two $q^2$ regions may indicate similar underlying physical mechanisms. These behaviors can be qualitatively understood as arising from the dominant effects of the charged Higgs sector, with additional contributions from the chargino sector appearing in $\mathrm{BR}(B \to X_{\mathrm{s}} l^{+} l^{-})_{\mathrm{high}}$.

To further elucidate the origin of these behaviors, it is instructive to examine the underlying effective operators. The dominance observed at the particle level translates into distinct new physics corrections to the relevant Wilson coefficients, as specified in Table~\ref{tab:dominant_contributions}. From the numerical analysis, we find that $\mathrm{BR}(B \to X_{\mathrm{s}} l^{+} l^{-})_{\mathrm{low}}$ are mainly governed by $C_{7}$, with the charged-Higgs sector providing the dominant contribution through $\gamma$-penguin diagrams. In contrast, $\mathrm{BR}(B \to X_{\mathrm{s}} l^{+} l^{-})_{\mathrm{high}}$ are dominated by $C_{9}$ and $C_{10}$, with these coefficients being predominantly influenced by $Z$-penguin diagrams. Only mild corrections stem from $\gamma$-penguin and box diagrams, such as $C_{9,\chi^\pm}^\gamma(\mu_{EW})$, $C_{9,S^\pm}^\gamma(\mu_{EW})$, $C_{9,\chi^\pm}^{box}(\mu_{EW})$, and $C_{10,S^\pm}^{box}(\mu_{EW})$. The resulting interplay among these contributions leads to a modest partial compensation, producing a slightly modified yet phenomenologically relevant total contribution. This conclusion aligns with the analysis in Refs.~\cite{Altmannshofer19,Altmannshofer:JHEP12}, where it is demonstrated that the angular observables $S_6^{s}$ and $S_6^{c}$ receive contributions from the Wilson coefficients $C_7$, $C_9$, $C_{10}$, and $C_S$. These contributions form the basis for understanding the observed $B \to X_{\mathrm{s}} l^{+} l^{-}$ and $A_{\mathrm{FB}}(B \to X_{\mathrm{s}} l^{+} l^{-})$. Furthermore, we now shift our focus to how these Wilson coefficients influence $A_{\mathrm{FB}}(B \to X_{\mathrm{s}} l^{+} l^{-})$, where the interference terms between the coefficients lead to significant and interesting effects.
\begin{figure}[t]
  \centering
  \includegraphics[width=\columnwidth]{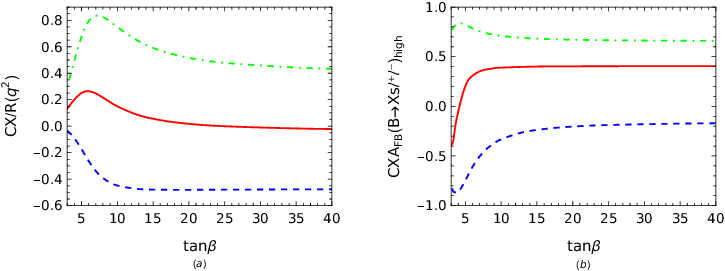}
   \caption{Decomposition of (a) $A_{\mathrm{FB}}(B \to X_{\mathrm{s}} l^{+} l^{-})_{\mathrm{low}}$ and (b) $A_{\mathrm{FB}}(B \to X_{\mathrm{s}} l^{+} l^{-})_{\mathrm{high}}$. Total $A_{\mathrm{FB}}$ (solid line), $C_7C_{10}$ interference (dashed line), and $C_9C_{10}$ interference(dash-dotted line).
}
\label{fig8}
\end{figure}

Eq.~(\ref{eq:normalized-AFB}) shows that the forward-backward asymmetry $A_{\mathrm{FB}}$ for $B \to X_{\mathrm{s}} l^{+} l^{-}$ is determined by the ratio of $\bar{A}_{\mathrm{FB}}$ to $R(q^2)$. As indicated by Eq.~(\ref{eq:Unnormalized-AFB}), $\bar{A}_{\mathrm{FB}}$ receives contributions from four interference terms, with the dominant ones identified as $C_7C_{10}$ and $C_9 C_{10}$ in our analysis. Since $R(q^2)>0$ over the kinematic range of interest, the behavior of $A_{\mathrm{FB}}(B \to X_{\mathrm{s}} l^{+} l^{-})$ is governed by the numerator $\bar{A}_{\mathrm{FB}}$. We therefore analyze $\bar{A}_{\mathrm{FB}}$ in terms of these interference structures. In the low-$q^2$ region, the contributions proportional to $C_7C_S$ and $C_9C_S$ are numerically negligible. The behavior of $A_{\mathrm{FB}}(B \to X_{\mathrm{s}} l^{+} l^{-})_{\mathrm{low}}$ as a function of $\tan\beta$ is governed by the relative importance of the $C_7C_{10}$ and $C_9C_{10}$ terms, leading to two distinct changes in the dominant contribution across different parameter ranges. As we concentrate on the region with $\tan\beta < 5.5$ in  Fig.~\ref{fig4:sixplots}(a) , $A_{\mathrm{FB}}(B \to X_{\mathrm{s}} l^{+} l^{-})_{\mathrm{low}}$ shows a consistent trend of increase as $\tan\beta$ increases. For $\tan\beta > 5.5$, this quantity decreases as $\tan\beta$ increases. This stems from a competition between four SUSY contributions.

To quantify the dominant interference contributions to $A_{\mathrm{FB}}(B \to X_{\mathrm{s}} l^{+} l^{-})$ as a function of $\tan\beta$, Fig.~\ref{fig8}(a) compares the total $A_{\mathrm{FB}}(B \to X_{\mathrm{s}} l^{+} l^{-})_{\mathrm{low}}$ with the individual contributions from the $C_7C_{10}$ and $C_9C_{10}$ interference terms. In the regime $\tan\beta < 5.5$, the total $A_{\mathrm{FB}}(B \to X_{\mathrm{s}} l^{+} l^{-})_{\mathrm{low}}$ follows the behavior of the $C_7C_{10}$ contribution, with the shifted peak reflecting the interplay and partial cancellation between the two dominant interference terms. As $\tan\beta$ increases and lies in the range $5.5 < \tan\beta < 7$, the dominant contribution gradually transitions from $C_7C_{10}$ to $C_9C_{10}$. For $\tan\beta > 7$, the $C_7C_{10}$ term becomes dominant again. Over the full $\tan\beta$ range, the combined $C_7C_{10}$ and $C_9C_{10}$ contributions reproduce the total $A_{\mathrm{FB}}$, confirming their quantitative dominance in the low-$q^2$ region.

A similar dominance transition persists in the high-$q^2$ region, although the different $q^2$ dependence of $A_{\mathrm{FB}}(B \to X_{\mathrm{s}} l^{+} l^{-})$ leads to a transition pattern distinct from that in the low-$q^2$ region. In the high-$q^2$ region, we systematically decompose the Wilson coefficient contributions to the angular coefficients governing $A_{\mathrm{FB}}(B \to X_s \ell^+ \ell^-)$. Since $A_{\mathrm{FB}}(B \to X_{\mathrm{s}} l^{+} l^{-})_{\mathrm{high}}$ is proportional to the $I_6^{s}$ term, we analyze how the individual Wilson coefficients enter the corresponding $A_{\perp L,R}$ and $A_{\parallel L,R}$-related contributions. Detailed definitions of these quantities are given in Ref.~\cite{Altmannshofer19}. The numerical results in Fig.~\ref{fig8}(b) exhibit a clear transition pattern: for $\tan\beta < 8$, the total $A_{\mathrm{FB}}$ follows the behavior of the $C_7C_{10}$ contribution, while for $\tan\beta > 8$, the $C_9C_{10}$ term becomes dominant and the total $A_{\mathrm{FB}}$ for $B \to X_{\mathrm{s}} l^{+} l^{-}$ shows only mild dependence on $\tan\beta$. The combined contributions from $C_7C_{10}$ and $C_9C_{10}$ reproduce the total $A_{\mathrm{FB}}$ behavior throughout the high-$q^2$ region.

To gain deeper insight into the mechanisms behind these effects, we now examine the contributions from both the relevant loops and the structure of the mixing matrices. Regarding the chargino-squark loop contributions to the Wilson coefficients $C_7$, $C_9$, $C_{10}$ and $C_S$, the possible enhancement originates from the chiral structure of the relevant interaction vertices. Following the discussion in Refs.~\cite{Eberl:PRD104,Carena:NPB577}, the $b - {u}_{i} - {\chi}_k$ and $s -{u}_{i} -{\chi}_k$ vertices involve couplings proportional to the bottom and top Yukawa couplings, $Y_b$ and $Y_t$, respectively, which can be enhanced in the large  $\tan\beta$ regime. Similarly, in the charged Higgs sector, Yukawa-enhanced contributions emerge and constitute the dominant source of corrections to the relevant Wilson coefficients. The Yukawa couplings of charginos (and similarly for neutralinos, which contribute even less) are directly related to their mass matrices. The $\tan\beta$-enhanced contributions from different eigenstates are suppressed by products of small CKM matrix elements, which partially compensate each other, resulting in a modest net effect on the Wilson coefficients. This partial compensation naturally explains the mild overall $\tan\beta$ dependence observed in these sectors. This behavior originates from the crucial dependence of the chargino (and neutralino) mass matrix elements on $\tan\beta$ via the vacuum expectation values $\upsilon_d$ and $\upsilon_u$. This induces subtle corrections to the Wilson coefficients, thereby modifying the theoretical predictions for physical observables. The explicit forms of the mass matrices and the corresponding diagonalization procedures can be found in Ref.~\cite{Zhang:NPB13} and related works, and are not repeated here. Moreover, relevant corrections to both the b-quark and t-quark Yukawa couplings in the large $\tan\beta$ regime lead to important contributions to $b$ to $s$ processes , which are qualitatively consistent with our results~\cite{Gorbahn:PRD84, Degrassi:JHEP12,Carena:PLB499,Bobeth:NPB630}.

\begin{figure}[t]
  \centering
  \includegraphics[width=\columnwidth]{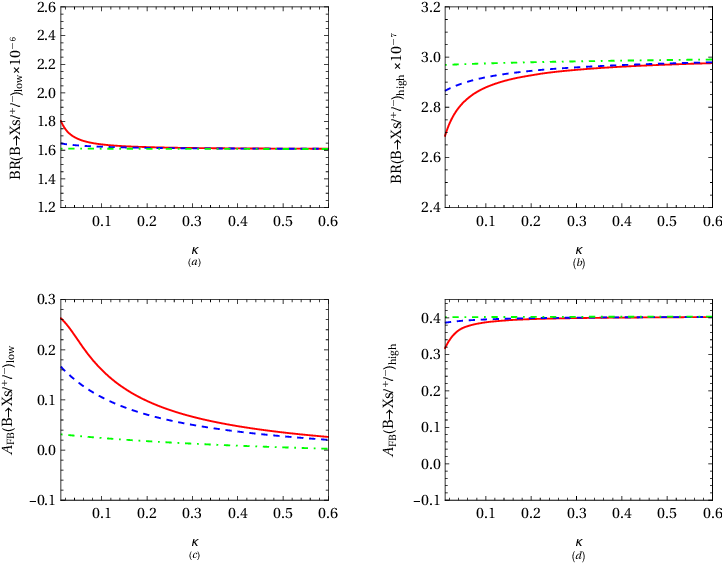}
   \caption{Fixed $\tan\beta=25$, $\upsilon_{\nu^c}=2.0~\mathrm{TeV}$:
(a) $\mathrm{BR}(B \to X_{\mathrm{s}} l^{+} l^{-})_{q^2 \in [1,6]~\mathrm{GeV}^2}$,
(b) $\mathrm{BR}(B \to X_{\mathrm{s}} l^{+} l^{-})_{q^2 \in [14.4,25]~\mathrm{GeV}^2}$,
(c) $\mathrm{BR}(B \to X_{\mathrm{s}} l^{+} l^{-})_{q^2 \in [1,6]~\mathrm{GeV}^2}$, and
(d) $\mathrm{BR}(B \to X_{\mathrm{s}} l^{+} l^{-})_{q^2 \in [14.4,25]~\mathrm{GeV}^2}$
vary with $\kappa$ for
$A_\lambda = 0.05~\mathrm{TeV}$ (solid line),
$A_\lambda = 0.2~\mathrm{TeV}$ (dashed line), and $A_\lambda = 1~\mathrm{TeV}$ (dashed-dotted line).
}
  \label{fig6:sixplots}
\end{figure}
\begin{figure}[t]
  \centering
  \includegraphics[width=\columnwidth]{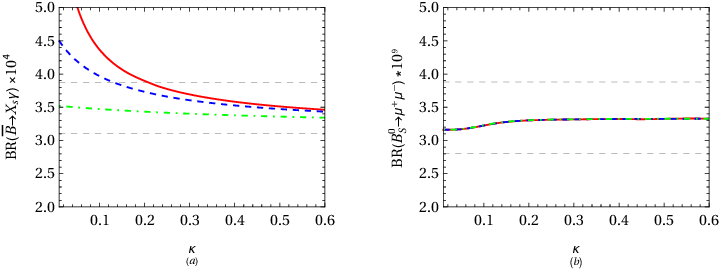}
   \caption{ Fixed $\tan\beta=25$, $\upsilon_{\nu^c}=2.0~\mathrm{TeV}$: (a) $\mathrm{BR}(\bar{B} \to X_{\mathrm{s}}\gamma)$  and (b) $\mathrm{BR}(B_{\mathrm{s}}^{0} \to \mu^{+} \mu^{-})$ vary with $\kappa$ for
$A_\lambda = 0.05~\mathrm{TeV}$ (solid line),
$A_\lambda = 0.2~\mathrm{TeV}$ (dashed line), and $A_\lambda = 1~\mathrm{TeV}$ (dashed-dotted line).
}
 \label{fig7:sixplots}
\end{figure}

To further analyze the influence of relevant parameters on $B \to X_{\mathrm{s}} l^{+} l^{-}$ decay, we investigate whether the same underlying physical mechanism governs the decay across different regions of the model parameter space. Fig. \ref{fig6:sixplots} shows $\mathrm{BR}(B \to X_{\mathrm{s}} l^{+} l^{-})$ and $A_{\mathrm{FB}}(B \to X_{\mathrm{s}} l^{+} l^{-})$ as functions of $\kappa$ for $\tan\beta=25$, while taking into account the experimental constraints from $\bar{B} \to X_{\mathrm{s}}\gamma$ and $B_{\mathrm{s}}^{0} \to \mu^{+} \mu^{-}$ shown in Fig. \ref{fig7:sixplots}. As can be seen from Eq.~\eqref{eq:MHc1}, in addition to $\tan\beta$, the parameter $\kappa$ also contributes to the charged Higgs boson mass. Similar to the extensions of the Standard Model discussed in Ref.~\cite{2006PRD74}, the new physics contributions to $B \to X_{\mathrm{s}} l^{+} l^{-}$ in the $\mu\nu$SSM primarily depend on $M_{H^{\pm}}$ and $\tan\beta$.  Therefore, $\kappa$ may also affect the predicted values of $\mathrm{BR}(B \to X_{\mathrm{s}} l^{+} l^{-})$ and $A_{\mathrm{FB}}(B \to X_{\mathrm{s}} l^{+} l^{-})$.

From Fig. \ref{fig6:sixplots}, we observe that the branching ratios of $B \to X_{\mathrm{s}} l^{+} l^{-}$ exhibit only mild variations with the parameter $\kappa$ in both the low- and high-$q^2$ regions, while their overall trends remain consistent with the results observed in the $\tan\beta$ scan. Across the entire $\kappa$ range, the charged Higgs contribution remains dominant. Variations in $\kappa$  mainly affect the overall magnitude of the observables, while preserving the relative hierarchy of contributions. $A_{\mathrm{FB}}(B \to X_{\mathrm{s}} l^{+} l^{-})$ shows a similar dependence on $\kappa$. In contrast to $\tan\beta$ scans, where the dominant interference can shift from $C_7C_{10}$ to $C_9C_{10}$ as $\tan\beta$ increases, variations in $\kappa$ do not induce a qualitative change in the dominant interference pattern. The leading contributions remain governed by $C_7C_{10}$ and $C_9C_{10}$, with only a redistribution of their relative weights.

The dominant interference pattern remains stable across the entire $\kappa$ range, resulting from partial compensations among the interference terms as their relative magnitudes are modulated by the $\kappa$ dependence of the Wilson coefficients. Variations in $\kappa$ primarily rescale the overall magnitude of the Wilson coefficients without altering their relative hierarchy or interference structure. Consequently, the physical interpretation from prior $\tan\beta$ analyses remains valid, demonstrating the robustness of the mechanism against variations in model parameters.
\section{Discussion on Additional Constraints\label{sec6}}
Although this work mainly focuses on the underlying mechanisms of the $b\to s$ transitions within the $\mu\nu$SSM framework, it is important to ensure that the adopted parameter regions are consistent with other low-energy observables. We therefore briefly comment on several additional phenomenological constraints relevant to the present R-parity-violating scenario.

The $\mu\nu$SSM can induce additional contributions to $B_s$-$\bar{B}_s$ mixing through chargino-, neutralino-, and scalar-mediated loop diagrams. However, unlike generic trilinear RPV scenarios, the origin of R-parity violation in the $\mu\nu$SSM is closely related to the last three terms in the superpotential (Eq.~\eqref{eq:superpotential}),
\begin{equation}
W \supset Y^\nu_{ij}\,\hat H_u \hat L_i \hat \nu^c_j
+\lambda_i\,\hat \nu^c_i \hat H_d \hat H_u + \frac{1}{3} \kappa_{ijk} \hat{\nu}_i^c \hat{\nu}_j^c \hat{\nu}_k^c.
\end{equation}
In the present analysis, the neutrino Yukawa couplings $Y^\nu_{ij}$ are negligibly small and are set to zero in the numerical evaluation,  which leads to suppressed lepton-number- and flavor-violating effects. Moreover, the $\lambda_i \hat{\nu}_i^c \hat H_d \hat H_u$ and $\kappa_{ijk}\hat{\nu}_i^c \hat{\nu}_j^c \hat{\nu}_k^c$ interactions mainly affect the singlet sneutrino, Higgsino, chargino, and neutralino sectors rather than directly inducing flavor-changing interactions in the quark sector.
As a result, their effects on $B_s$--$\bar{B}_s$ mixing can only arise indirectly through chargino-, neutralino-, and scalar-mediated loop corrections. Within the allowed parameter space of the present work, these contributions are further suppressed by the small left-handed sneutrino vacuum expectation values and the TeV-scale supersymmetric particle spectrum.
In addition, the benchmark parameter regions considered here do not involve sizable off-diagonal soft flavor structures. Consequently, the resulting corrections to $\Delta M_s$ and the mixing phase $\phi_s$ remain compatible with current experimental bounds.

Exclusive $B\to K^*\mu^+\mu^-$ and $B\to K^*\nu\bar{\nu}$ observables also provide important constraints on the relevant Wilson coefficients. Within the parameter space considered in this work, the predicted branching ratios and forward-backward asymmetries are consistent with current experimental data. A more detailed study of these observables in the $\mu\nu$SSM framework will be presented elsewhere.

Recently, global analyses of $b\to s$ transitions~\cite{Hurth:JHEP1410,Altmannshofer:EPJC382,Aebischer:EPJC19,Alguero:EPJC82} have identified the allowed regions in the space of the new physics Wilson coefficients $C_{7,NP}$, $C_{9,NP}$, and $C_{10,NP}$ consistent with current experimental data. Within the allowed parameter space of the present work, the new physics contributions of the $\mu\nu$SSM to the relevant Wilson coefficients remain consistent with the experimentally measured branching ratio of $B\to X_s\mu^+\mu^-$ and lie within the corresponding $68\%$ confidence-level preferred regions reported in Ref.~\cite{Hurth:JHEP1410}.

\section{sumarry\label{sec7}}
In this paper, we investigate the branching ratios and forward-backward asymmetries of the rare decay $B \to X_{\mathrm{s}} l^{+} l^{-}$ in the $\mu\nu$SSM, incorporating the experimental constraints from $\bar{B} \to X_{\mathrm{s}}\gamma$ and $B_{\mathrm{s}}^{0} \to \mu^{+} \mu^{-}$. We focus on the impact of new physics on these decays through corrections to the Wilson coefficients at the hadronic scale. Within the $\mu\nu$SSM, we evaluate the relevant Wilson coefficients and identify $C_7$, $C_9$, $C_{10}$ and $C_S$ as the dominant sources of new-physics effects in the decays $\bar{B} \to X_{\mathrm{s}}\gamma$, $B_{\mathrm{s}}^{0} \to \mu^{+} \mu^{-}$, and $B \to X_{\mathrm{s}} l^{+} l^{-}$.

A systematic interference decomposition of the Wilson-coefficient contributions to the forward-backward asymmetry shows that its behavior is governed predominantly by the $C_7C_{10}$ and $C_9C_{10}$ interference terms, whose interplay gives rise to the distinct behaviors observed in the low- and high-$q^2$ regions.

The new-physics contributions to $B \to X_{\mathrm{s}} l^{+} l^{-}$ depend primarily on the charged-Higgs mass $M_{H^{\pm}}$ and the parameter $\tan\beta$, mainly through the enhancement of the Yukawa couplings. Partial cancellations among the chargino and neutralino mixing matrices play only a subleading role. Moreover, the parameters $\upsilon_{\nu^{c}}$, $\kappa$, and $A_{\lambda}$ have only a minor impact on the theoretical predictions for $B \to X_{\mathrm{s}} l^{+} l^{-}$.

\begin{acknowledgments}

The work has been supported by the National Natural Science Foundation of China (NNSFC) with Grants No. 12235008, Hebei Natural Science Foundation with Grants No. A2023201041.
\end{acknowledgments}

\end{document}